\documentclass{iopart}
\usepackage{amssymb,booktabs,color,url,verbatim}
\usepackage{graphicx}
\usepackage[colorlinks=true,citecolor=black,filecolor=black,linkcolor=black,urlcolor=black,pdftex]{hyperref}
\newcommand{\arxiv}[1]{\href{http://arxiv.org/abs/#1}{\texttt{arXiv:#1}}}
\newcommand{\1}[1]{\, \mathrm{#1}} 
\newcommand{\n}[1]{\mathrm{#1}}    
\newcommand{\degree}{{}^{\circ}}
\newcommand{\degrees}{\degree}
\newcommand{\percent}{\%}
\newcommand{\order}{\mathcal{O}}
\newcounter{fff}\renewcommand{\thefff}{({\footnotesize \Alph{fff}})}
\newcommand{\feature}[1]{\refstepcounter{fff}\thefff~\textit{#1}}
\definecolor{orange}{rgb}{1,0.5,0}
\begin{document}
\title{Search for Dark Matter with CRESST}
\author{Rafael F. Lang, Wolfgang Seidel}
\address{Max-Planck-Institut f\"ur Physik, F\"ohringer Ring 6, D-80805 M\"unchen, Germany}
\ead{rafael.lang@mpp.mpg.de}
\begin{abstract}
The search for direct interactions of dark matter particles remains one of the most pressing challenges of contemporary experimental physics. A variety of different approaches is required to probe the available parameter space and to meet the technological challenges. Here, we review the experimental efforts towards the detection of direct dark matter interactions using scintillating crystals at cryogenic temperatures. We outline the ideas behind these detectors and describe the principles of their operation. Recent developments are summarized and various results from the search for rare processes are presented. In the search for direct dark matter interactions, the CRESST-II experiment delivers competitive limits, with a sensitivity below $5\times10^{-7}\1{pb}$ on the coherent WIMP-nucleon cross section.

\end{abstract}
\submitto{\NJP}
\maketitle


\section{Introduction}

The overwhelming evidence for dark matter is discussed in great detail in other contributions to this \textit{Focus Issue}~\cite{clowe2009,kuhlen2009,jedamzik2009}. Particularly well motivated candidates for dark matter are Weakly Interacting Massive Particles, WIMPs for short~\cite{bergstrom2009,friedland2009,matchev2009}. On the one hand, the experimental particle physicist tries to produce such particles at colliders~\cite{baer2009,battaglia2009,hewett2009}. The observational astrophysicist on the other hand tries to detect their signature from Space, by recording their annihilation products~\cite{serpico2009,conrad2009,hailey2009,picozza2009}. Yet a third discovery channel exists through the identification of direct interactions of WIMPs with a target.

Today, this third detection channel is limited by the technological possibilities, which are pushed to their limits. A variety of experiments, each a distinct development, is set up to discover WIMPs through their direct interaction~\cite{collar2009,kim2009,nikkel2009,mohapatra2009,oberlack2009,sciolla2009}. In this \textit{Focus Issue} contribution, we review the search using scintillating cryogenic calorimeters and present recent advances. This technology is pursued by the CRESST~\cite{angloher2009} and ROSEBUD~\cite{cebrian2004} collaborations and will be used for the future EURECA detector~\cite{kraus2009b}.

To start, let us define the requirements imposed on any experiment that takes up the challenge to directly detect interactions induced by WIMPs. For direct WIMP-nucleus scattering, the detection rate
\begin{eqnarray}
	\Gamma = n_{\n{target}} \Phi \, \sigma'\label{eq:simplerate}
\end{eqnarray}
(where $n_{\n{target}}$ is the number of target atoms, $\Phi$ the flux of WIMPs, and $\sigma'$ the cross section of the process) is obviously very low. From this one can already deduce two major requirements for any experiment that aims to detect such interactions: \feature{it needs to provide a large target mass}\label{req:large} and \feature{it needs to be operational for a long time}\label{req:long}, so that collected exposures account for many kilogram$\times$weeks.

Spin-independent WIMP interactions can take place coherently over the full target nucleus. Then all the individual WIMP-nucleon scattering amplitudes $\sigma$ add up in phase, increasing the above detection rate (\ref{eq:simplerate}) to
\begin{eqnarray}
	\Gamma = n_{\n{target}} \Phi \, \sigma A^2\label{eq:coherentrate}
\end{eqnarray}
where $A$ is the mass number of the target material. Therefore, it is advantageous for direct searches to use \feature{heavy target nuclei}\label{req:heavy} to search for spin-independent WIMP interactions~\cite{cerdeno2009}.

The maximum recoil energy $E_{\n{r,max}}$ can readily be estimated to be
\begin{eqnarray}
E_{\n{r,max}} = \frac{(2p_{\chi})^2}{2m_{\n{N}}}  
          \sim \frac{\left( 2\times100\1{GeV/c^2}\times10^{-3}\1{c} \right)^2}{2\times100\1{GeV/c^2}} = 200\1{keV}
\end{eqnarray}
where $m_{\n{N}}$ is the target nucleus' mass and $p_{\chi}$ the WIMP momentum in the laboratory frame. Here, we assumed a typical WIMP with mass $m_{\chi}\approx100\1{GeV/c^2}$ expected from extensions of the Standard Model of particle physics~\cite{bertone2005,jungman1996}, and an average velocity $\langle v \rangle\approx10^{-3}\1{c}$ typical for Galactic velocities in our neighborhood~\cite{kerr1986,donato1998}. However, $E_{\n{r,max}}$ is an upper limit, and typical recoil energies are much lower, as we will see shortly. Hence, a major challenge for such experiments are \feature{an energy threshold low enough given the small energies expected from an WIMP interaction}\label{req:threshold}, and \feature{a way to calibrate the nuclear recoil energy scale in this range}\label{req:calibration}.

Properties of the phase space distribution of dark matter in our Milky Way and at the position of the Earth are inferred from N-body simulations~\cite{vogelsberger2009,kuhlen2009}. As an approximation to the true distribution, the isothermal halo~\cite{binney1994} is commonly used due to its simplicity and since it reproduces the observed flat rotation curve~\cite{honma1997}. Hence, the prototypical velocity distribution for WIMPs is the Maxwell-Boltzmann-distribution $f_{\n{MB}}$. In this first approximation, the expected spectrum of WIMP induced nuclear recoils is a simple exponential:
\begin{eqnarray}
\frac{\n{d}\Gamma}{\n{d}E_{\n{r}}} \propto \; \Phi \; \propto \; \langle v \rangle \; \propto \; \int_{v_{\chi}}^{\infty} f_{\n{MB}}(v)\; v \;\n{d} v \; \propto \n{e}^{-c_1v_{\chi}^2}  \; \propto \; \n{e}^{-c_2E_r}\label{eq:dnde}
\end{eqnarray}
where $v_{\chi}=\sqrt{2E_{\n{r}}/m_{\n{N}}}$ is the minimum WIMP velocity causing a recoil of energy $E_{\n{r}}$, and $c_{1,2}$ are positive constants to make the exponential dimensionless. Therefore, no line signature is expected in direct WIMP scattering searches, and this leads to additional requirements for these experiments. Most importantly, very efficient ways to discriminate the signal from various backgrounds are mandatory: Experiments \feature{need to be shielded against ambient radioactivity}\label{req:shielding}, need to be placed deep underground to reduce the background induced by cosmic rays, and care must be taken in the selection of materials used in the vicinity of the detector. Beyond this passive shielding, experience shows that \feature{active discrimination between signal and background}\label{req:discrimination} is required. In particular, one needs to have some means to \feature{distinguish neutron induced nuclear recoils from WIMP induced ones}\label{req:neutrons}. Information on the interaction coordinates can help to reduce the background, and of course, a directional sensitivity would be a good way to distinguish signal from background.

Efforts to understand background processes at the energies of interest are showing large improvements in recent years. In case of a signal in the experiment, systematic effects and many radioactive backgrounds can be excluded as alternative explanations using different target materials. Given current technology, this \feature{multi-target approach}\label{req:multitarget} seems a necessary requirement for the community to accept a claim of the detection of a WIMP induced signal. 

As long as no discovery is claimed, an upper limit on the WIMP-nucleon cross section is calculated. Due to the quasi exponential form of the spectrum, one would like to consider energies as close to threshold as possible. To do this, it is necessary to \feature{demonstrate the trigger efficiency}\label{req:trigger} over the full energy range used. 

\section{Cryogenic Calorimeters}

Most of the energy in an elastic scattering interaction is deposited in a target as phonons. Therefore, it is natural to meet the threshold challenge~\ref{req:threshold} with calorimetric devices, where the principal part of the deposited energy is converted into a signal.

In addition, such devices have the advantage that they can give a precise measurement of the deposited energy, as we will see shortly. Also, the excitation energy of phonons is low compared to that of charge carriers. Statistical fluctuations in the signal due to Poisson statistics are therefore minimized, which further reduces uncertainties in the energy inferred from an interaction.

\subsection{Superconducting Phase-Transition Thermometers}

As a first approximation, we can model the calorimeter as a heat capacity $C$, and assume the phonons generated in an interaction are in thermal equilibrium. Then, the increase of temperature $\Delta T$ in the calorimeter following a particle interaction is simply
\begin{eqnarray}
	\Delta T = \frac{\Delta E}{C},\label{eq:calorimeter}
\end{eqnarray}
where $\Delta E$ is the deposited energy. Since heat capacities decrease with decreasing temperature, cooling the calorimeter leads to a larger temperature signal for a given energy deposition. Commercially available $\n{{}^{3}He/{}^{4}He}$ dilution refrigerators can readily achieve temperatures in the millikelvin range and are used in many direct dark matter experiments. More generally, cryogenic calorimeters are a mature technology with applications in many different fields of science~\cite{enss2005,enss2008}.

\begin{figure}[htbp]
\begin{flushright}\includegraphics[width=0.8\columnwidth]{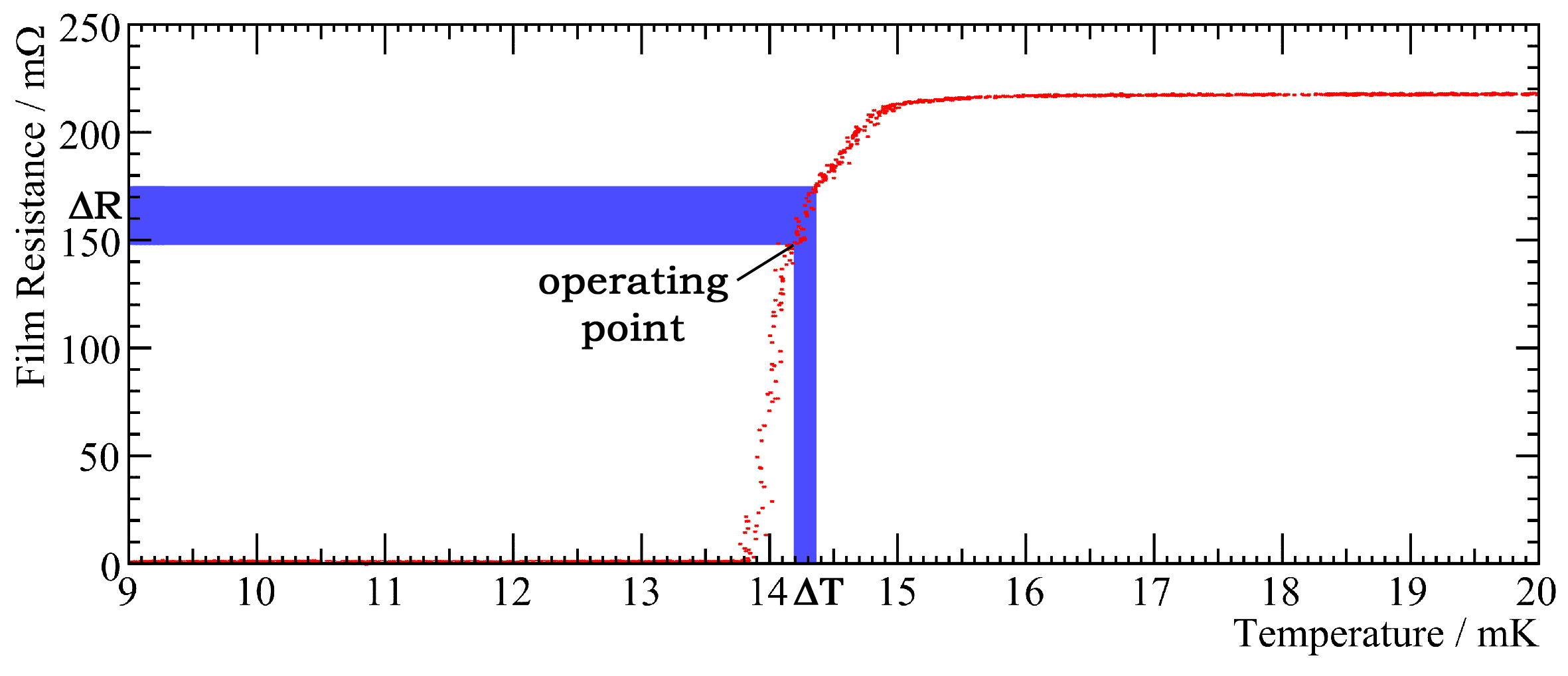}\end{flushright}
\vspace{-5mm}\caption{Operating principle of superconducting phase transition thermometers (SPTs). The thermometer is stabilized in its transition to the superconducting state, where its resistance $R$ is strongly varying with temperature $T$. This allows to measure the increase of resistance $\Delta R$ following a particle interaction, hence deducing its energy.}
\label{fig:transition}\end{figure}

Thermometers with a measurable change of resistance at millikelvin temperatures are superconducting phase transition thermometers (SPTs), in use for a variety of different applications~\cite{meier1999,cabrera2008}. These thin metal films are stabilized in their transition to the superconducting state, where the change of resistance $\Delta R$ given a change of temperature $\Delta T$ is large, see figure~\ref{fig:transition}. 

Materials that become superconducting in the millikelvin range are rare. Tungsten in its alpha phase becomes superconducting below $\sim15\1{mK}$~\cite{gibson1964} and can be used directly as thermometer film~\cite{colling1993}. Monocrystalline iridium has a transition temperature of $112\1{mK}$~\cite{hein1962,gubser1973}, too high for the purposes here, but it can be lowered and adjusted using the proximity effect~\cite{werthamer1963} in an iridium-gold bilayer~\cite{degennes1964,nagel1994}.

A typical thermometer used in the CRESST-II experiment as a phonon detector on a target crystal is a $200\1{nm}$ thick $6\times8\1{mm^2}$ tungsten film~\cite{angloher2004}. For metal films, the electronic contribution to the specific heat is $\gamma \, T$ with $\gamma\approx1\1{mJ\,mol^{-1}\,K^{-2}}$ for tungsten~\cite{kittel2005}. Hence, a particle depositing $10\1{keV}$ at an operating temperature of $20\1{mK}$ gives a temperature rise of $80\1{\mu K}$, assuming a heat capacity as for a normal conducting metal film.

To read out the corresponding small change of resistance of the low-ohmic thermometer, a SQUID based readout is used~\cite{seidel1990}. Superconducting Quantum Interference Devices are extremely sensitive in measuring the magnetic flux of an input coil. A bias current is split in two branches, one through the thermometer, the other through reference resistors and the input coil of the SQUID. This scheme allows to read out the temperature of the thermometer with a precision of a few $\mu K$. As a consequence, energy thresholds well below $1\1{keV}$ are readily obtained with such devices~\cite{angloher2002,guetlein2008}. Details of the SQUID system and data acquisition electronics used in CRESST-II can be found in~\cite{henry2007}.

\subsection{Model of Pulse Formation}\label{sec:phonons}

The thermodynamical picture of equation~(\ref{eq:calorimeter}), $\Delta T = \Delta E/C$, needs to be refined in order to understand the process of pulse formation in superconducting phase transition thermometers. A more detailed model has been developed in~\cite{proebst1995}: An energy deposition in the absorber leads to a spectrum of high frequency optical phonons within less than a nanosecond. Electron recoils from ionizing radiation eventually result in an almost monoenergetic population of acoustic phonons with about half the Debye frequency $\nu_{\n{Debye}}$. On the other hand, nuclear recoils from elastic scatterings on nuclei result in a continuous spectrum of acoustic phonons up to $\nu_{\n{Debye}}$. 

For $\n{CaWO_4}$ as a target material, the Debye temperature is $\Theta_{\n{Debye}}=250\1{K}$~\cite{gluyas1973}, corresponding to $\nu_{\n{Debye}}/2=k_{\n{B}}\Theta_{\n{Debye}}/2h\approx2\1{THz}$. THz phonons have energies of a few $\n{meV}$, which is large compared to thermal energies of interest, $E=k_{\n{B}}T=8.6\times10^{-5}\1{eV/K} \times 15\1{mK}=1\1{\mu eV}$. Hence, such THz phonons are called non-thermal phonons. They decay due to crystal lattice anharmonicities with a decay rate that is proportional to $\nu^5$~\cite{orbach1964}, or scatter with a rate proportional only to $\nu^4$~\cite{maris1990}. At a few $100\1{GHz}$, still above thermal energies, and on a time scale of a few milliseconds, the phonons spread ballistically throughout the absorber and fill it uniformly after a few reflections on the surface.

\begin{figure}[htbp]
\begin{flushright}\includegraphics[width=0.8\columnwidth,clip,trim=0 590 150 50]{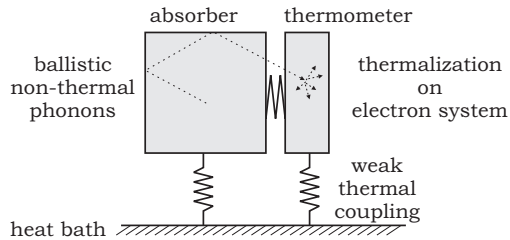}\end{flushright}
\vspace{-5mm}\caption{A particle interaction leads to non-thermal phonons that spread ballistically through the absorber. The thermometer has a strong thermal coupling to the absorber. Non-thermal phonons penetrate this coupling and are thermalized on the electron system of the superconducting phase transition thermometer.}
\label{fig:detectormodel}\end{figure}

Such non-thermal phonons are readily absorbed by the free electrons of the metal film in the thermometer (figure~\ref{fig:detectormodel}). This absorption is mediated by a strong coupling due to space charge variations of the phonon oscillation. The strong interaction among the electrons in the thermometer then quickly disperses and thus thermalizes the phonon energy, which results in a heating of the thermometer. With superconducting phase transition thermometers, this temperature change is then measurable as a change in resistance. Since phonons are mostly absorbed in the thermometer alone, this allows to scale up the targets to the large dimensions used in the CRESST-II experiment~\cite{ferger1994,loidl2001} and anticipated for future 1-ton-scale experiments, hence meeting the challenge~\ref{req:large} of a large target mass.

\subsection{Target Materials}

Challenge~\ref{req:discrimination}, being able to actively discriminate the nuclear recoil signal from the dominant electron recoil backgrounds, requires a second detection channel. In this \textit{Focus Issue} contribution, we consider as discrimination parameter the light produced by particles impinging on scintillating materials~\cite{gonzalez1989,alessandrello1992,bobin1997}. The target material needs to be coolable to millikelvin temperatures, and it needs to show a good scintillation efficiency at such low temperatures. In addition, the material needs to have a very low level of intrinsic radioactive contaminations. 

Many different materials are known to be good scintillators, yet few of them meet all these requirements, see~\cite{mikhailik2006} for a review of recent studies. The light yield of common doped scintillators such as NaI(Tl) or CsI(Tl) as well as rare-earth-doped scintillators such as $\n{CaF_2(Eu)}$ or $\n{YAlO_3(Ce)}$ is strongly degraded at cryogenic temperatures. Hence, self-activated scintillators are employed, since these have a high light yield even at low temperatures.

Most commonly, $\n{CaWO_4}$ crystals are used as a target~\cite{angloher2009,angloher2005}, since they have a high light yield even at cryogenic temperatures~\cite{mikhailik2007,moszynski2005} and were shown to be suitable for the task already ten years ago~\cite{meunier1999}. The dimensions of these target crystals is limited only by the Czochralski crystal growth process~\cite{brandle2004}, and optically clean $\n{CaWO_4}$ crystals with masses of $\sim1\1{kg}$ are commercially available. The crystals used in CRESST-II are cut in cylindrical shape with a diameter of $4\1{cm}$ and similar height, and weigh about $300\1{g}$ each.

Another target material recently receiving much attention is $\n{ZnWO_4}$~\cite{kraus2005,danevich2005}. Besides being suitable for the application at millikelvin temperatures~\cite{kraus2009}, it promises to be cleaner and available in larger specimens~\cite{nagornaya2008}. $\n{ZnWO_4}$ has a lower melting temperature of $\sim1200\degrees \n{C}$~\cite{grabmaier1984} in comparison to $\n{CaWO_4}$, which melts only at $\sim1600\degrees \n{C}$~\cite{nassau1962}. If the thermometer is evaporated directly onto the $\n{ZnWO_4}$ crystal, this changes the stoichiometry of the crystal, and hence its scintillation properties. Only recently it was shown that this problem can be overcome using a gluing technique~\cite{lanfranchi2004}. There, the thermometer is evaporated or sputtered~\cite{roth2008} on a small carrier crystal, which is subsequently glued to the large target crystal. Measurements show that non-thermal phonons penetrate the glue layer, as can be described by simple models~\cite{roth2009,kiefer2009}. Results from these investigations look very promising and enable the easy mass production of such detectors for future ton-scale dark matter searches. Thus, $\n{ZnWO_4}$ crystals with glued thermometers are already used in the search for dark matter~\cite{bavykina2008,bavykina2009a}.

$\n{CaMoO_4}$ is an interesting target material due to the strongly differing WIMP recoil spectrum ($A_{\n{Mo}}=96$), but otherwise similar structure, hence ideally meeting the multi target requirement~\ref{req:multitarget}. Although available specimen have shown inferior light yield at room temperature~\cite{bavykina2009a,annenkov2008}, the light yield increases towards lower temperatures, and eventually becomes comparable to that of $\n{CaWO_4}$~\cite{mikhailik2007b}. $\n{CdWO_4}$ may be used as an alternative target but shows a high intrinsic radioactivity~\cite{gironi2009}. Investigations using BGO~\cite{coron2009,calleja2008} and doped sapphire crystals~\cite{mikhailik2005,distefano2008,chapellier2009,luca2009} are under way with very promising results. All these different available materials under consideration make the multi-target requirement~\ref{req:multitarget} easy to meet.

\subsection{The Form Factor}

\begin{figure}[htbp]
\begin{flushright}\includegraphics[angle=90,width=0.8\columnwidth,clip,trim=0 0 140 0]{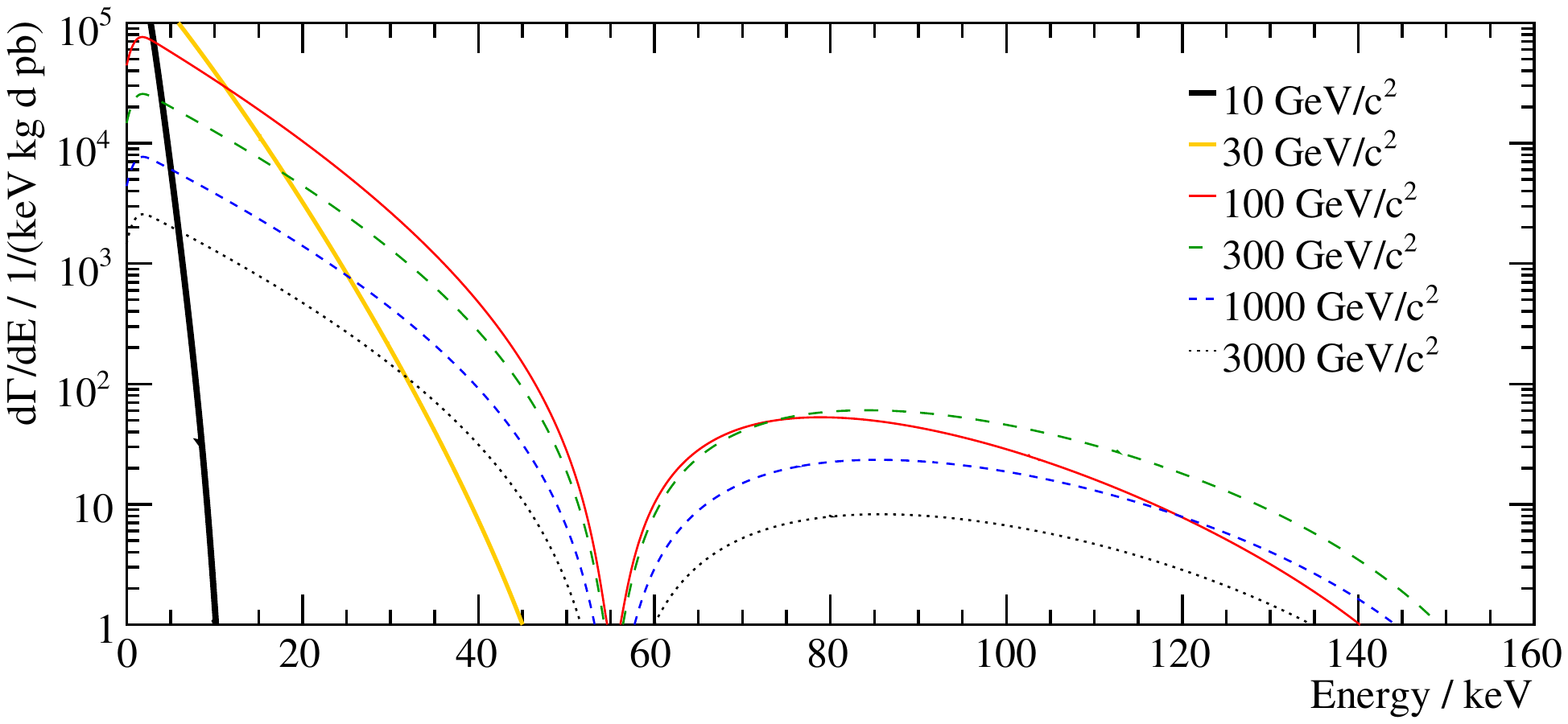}\end{flushright}
\vspace{-5mm}\caption{The expected WIMP recoil spectrum in a pure tungsten target for various WIMP masses, calculated for the standard WIMP scenario~\cite{donato1998}. Units are per $1\1{kg\,d}$ of exposure, for a WIMP-nucleon cross section of $1\1{pb}$, and per $1\1{keV}$ energy bin. For light WIMPs, the recoil spectrum is almost the simple exponential expected from equation~\ref{eq:dnde}. For heavy WIMPs, the spectral shape is completely dominated by the form factor, here assumed to be of Helm type.}
\label{fig:WdNdEC}\end{figure}

Tungsten ($A_{\n{W}}=184$) is the dominant scatter center (for experiments with a threshold energy below $\sim30\1{keV}$) for WIMPs in $\n{CaWO_4}$, due to the $\propto A^2$ dependence of the coherent WIMP interaction rate (equation~\ref{eq:coherentrate} and benefit~\ref{req:heavy}). This significantly enhances the sensitivity of experiments with tungsten targets. However, form factor effects become important. As is shown in figure~\ref{fig:WdNdEC}, the expected spectral shape of WIMP induced tungsten recoils is completely dominated by the form factor for WIMP masses above $m\approx 100\1{GeV/c^2}$, introducing a systematic uncertainty to be dealt with. As a parametrization for the form factor, customarily the Helm form factor~\cite{helm1956} is used since shown adequate by J.~Engel~\cite{engel1991}. Its parameters are adopted from data of electron scattering experiments, so an implicit assumption is that the WIMP scattering centers are distributed as the charge in the nucleus~\cite{lewin1996}, and alternative parameterizations~\cite{duda2007,olive2009} may be used instead. However, we find that the integrated scattering rate above a given energy threshold typically changes only at the percent level for most WIMP masses and form factor models.

\subsection{Light Detector}

To detect the scintillation light, the operation of photomultiplier tubes is not feasible, because at these low temperatures, the photocathode becomes non-conductive. In addition, photomultipliers are generally not radiopure, and the required high voltage causes problems in millikelvin applications. Hence, a dedicated cryogenic light detector is used in addition to the phonon detector.

\begin{figure}[htbp]
\begin{flushright}\includegraphics[width=0.8\columnwidth]{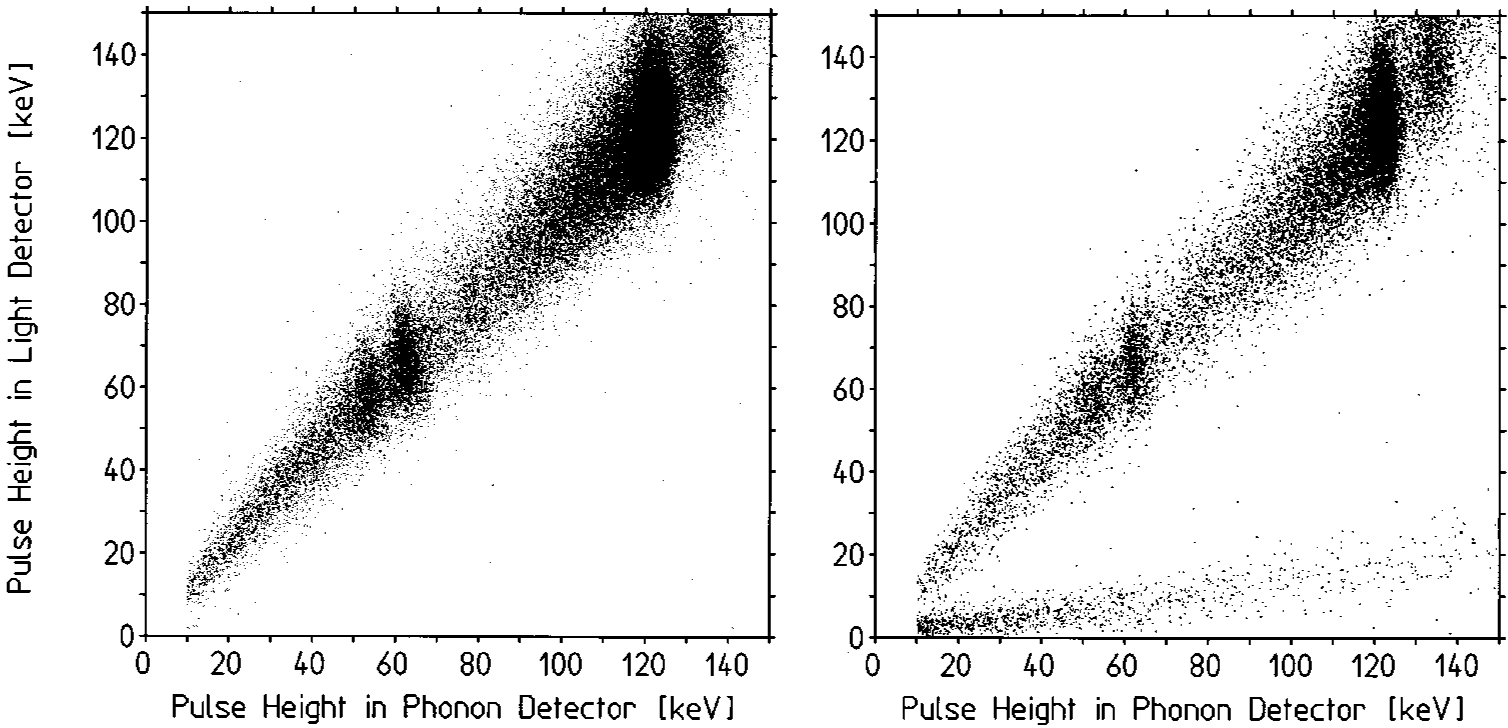}\end{flushright}
\vspace{-5mm}\caption{Operating a scintillating crystal as cryogenic target together with an appropriate light detector allows the discrimination of electron and nuclear recoils at the low energies relevant for a dark matter search. Shown is a scatter plot of the energy in the light detector versus the energy in the phonon detector, on the left with a $\n{{}^{57}Co}$ gamma source and a $\n{{}^{90}Sr}$ beta source, and on the right with an additional $\n{Am/Be}$ neutron source. Electron and gamma events are clearly separated from neutron induced events. In addition, one can see that electron and gamma events form one common band. This demonstrates the absence of any measurable light yield degradation near the surface of the crystal. Figure from~\cite{meunier1999}.}
\label{fig:proofofprinciple}\end{figure}

This light detector consists of a thin light absorbing substrate, equipped with another phase transition thermometer~\cite{petricca2004}. Following a particle interaction in the crystal, the scintillation light is absorbed in the wafer, creates phonons there, and can hence be detected with a similar phase transition thermometer. This allows to discriminate electron or gamma events from nuclear recoils~\cite{birks1967}, as is illustrated in figure~\ref{fig:proofofprinciple} showing data from the original proof-of-principle experiment.

This discrimination scheme has distinct advantages over other techniques: Degradation of the light signal for events that happen close to the crystal surface is not present~\cite{cozzini2004}, in contrast to ionizing detectors where surface effects may lead to a misidentification of an electron recoil as a nuclear recoil. This can also be seen in figure~\ref{fig:proofofprinciple}, since the gamma lines which originate throughout the crystal are in the same band as the continuous background from electrons scattering on the surface of the crystal. In addition, since only a few percent of the signal are emitted as light~\cite{westphal2006,frank2002b,distefano2003}, the quenching of the phonon channel can be neglected, giving a direct measurement of the energy from that channel. 

In CRESST-II, silicon-on-sapphire wafers are commonly used as light absorbing structure~\cite{petricca2004}, but superconducting materials are an interesting alternative~\cite{pantic2009}. The Neganov-Luke effect describes the enhancement of the light signal by the use of an electric field~\cite{neganov1985,luke1988}. This can be used to obtain an improved performance of the light detectors~\cite{isalia2008}, but stability issues have not yet been resolved to the precision required for the long exposures needed in a WIMP search.

\subsection{Detector Modules}\label{sec:modules}

\begin{figure}[htbp]
\begin{flushright}\includegraphics[width=0.8\columnwidth,clip,trim=0 590 158 0]{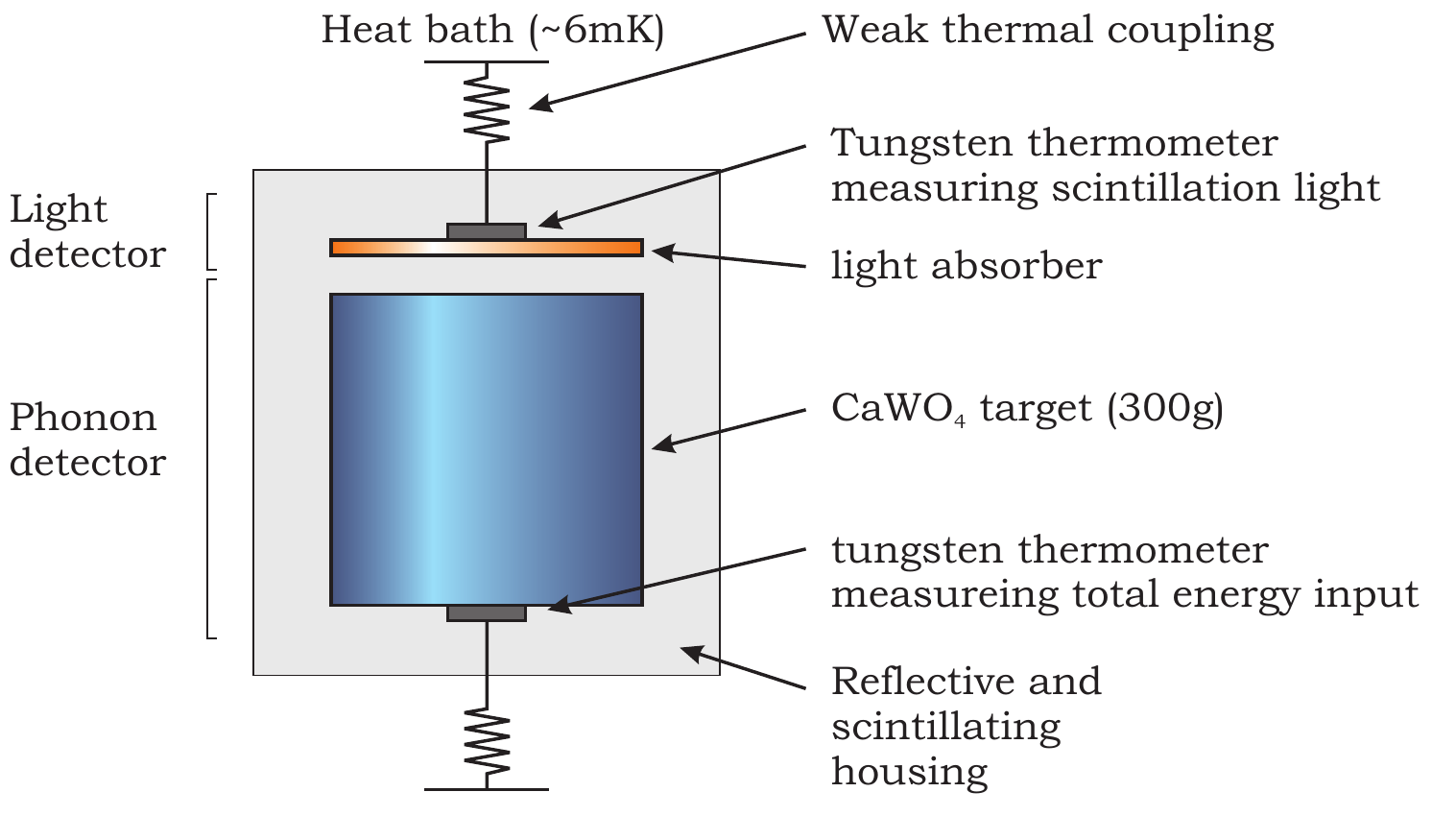}\end{flushright}
\vspace{-5mm}\caption{Concept of a CRESST-II detector module. Each module contains two separate tungsten thermometers. One is directly attached to the target crystal, measuring the total deposited energy. To measure the scintillation light, a thin light absorber together with the second thermometer is placed in the vicinity of the crystal. The structure is encapsulated in a reflective and scintillating housing.}
\label{fig:moduleschema}
\end{figure}

\begin{figure}[htbp]
\begin{flushright}\includegraphics[width=0.8\columnwidth]{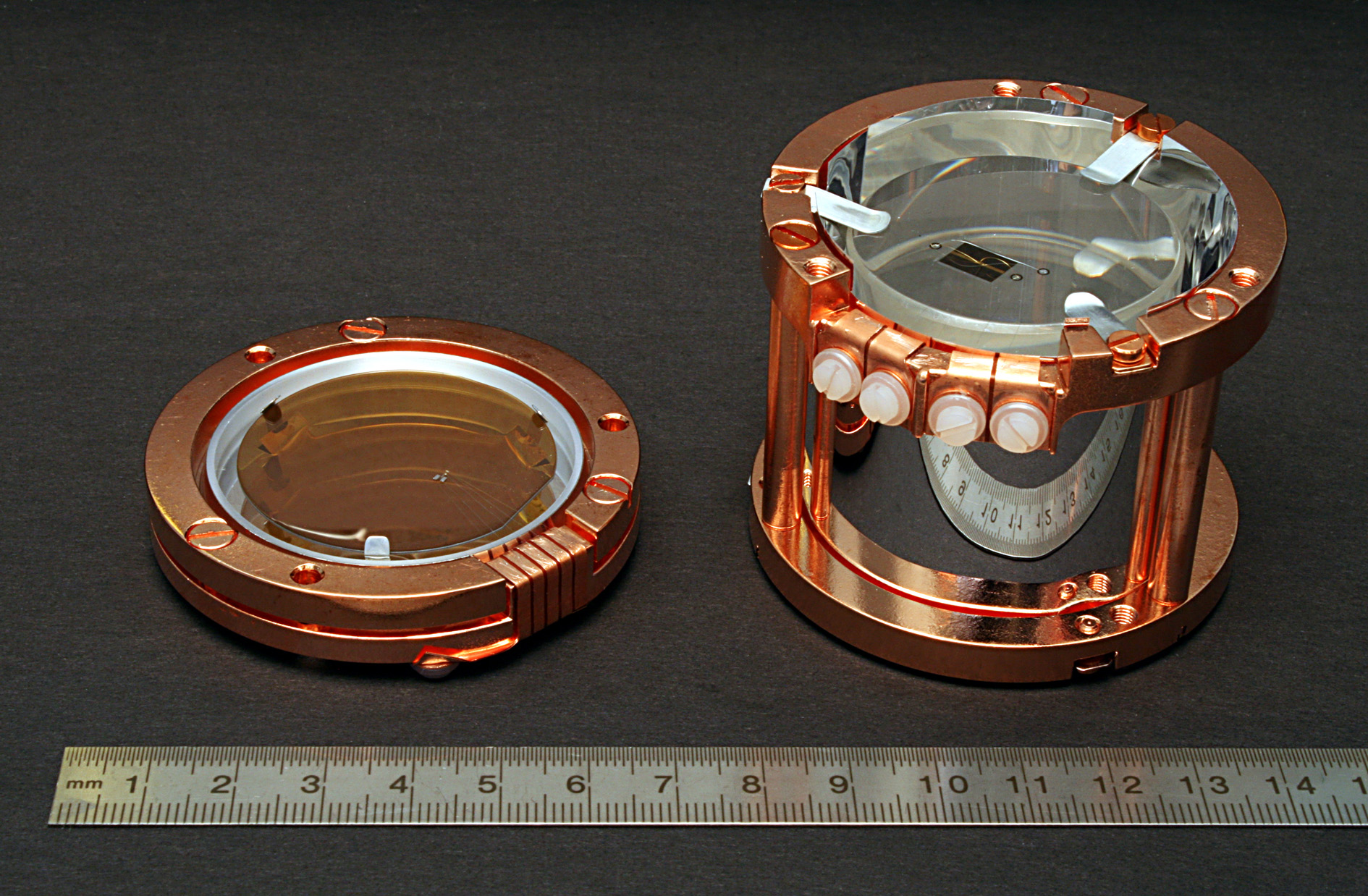}\end{flushright}
\vspace{-5mm}\caption{Picture of an opened detector module as used in CRESST-II. On the left, the (orange) wafer is the silicon-on-sapphire light absorber, the transition edge sensor is the tiny structure on it. On the right, the crystal can be seen with the phonon detector evaporated on it. The crystal is enclosed by the scintillating reflective foil and held by custom made springs. Bonding wires for thermal and electrical contact are just about visible.}
\label{fig:modulfotok}\end{figure}

The target crystal with its phonon detector and the light absorber with \textit{its} phonon detector are paired together in a housing and then referred to as a detector module. Figure~\ref{fig:moduleschema} illustrates the concept, and figure~\ref{fig:modulfotok} is a picture of an opened CRESST-II detector module. This modular structure makes these experiments almost trivially scalable to larger target masses, as dictated by the large mass requirement~\ref{req:large}.

To increase the amount of light collected by the light absorber, the module is encapsulated in a reflector. Due to the cylindrical geometry of the holder, light may undergo many reflections before it hits the light absorber, so a highly effective reflector is mandatory. Hence, the CRESST-II detectors are encapsulated in a Radiant Mirror Film VM2000/VM2002 from 3M~\cite{weber2000}. These polymeric foils achieve a high reflectivity through a multilayer structure which is optimized for light guide applications, but suitable also for the $\n{CaWO_4}$ emission spectrum~\cite{grasser1982}. 

A dangerous background for the search for dark matter comes from alpha decays in the vicinity of the crystal. If the alpha escapes, the recoiling daughter nucleus may impinge on the crystal and mimic a heavy nuclear recoil~\cite{westphal2008a,angloher2009}. In order to be able to distinguish such events from dark matter induced tungsten recoils, the reflective foil is also scintillating. Then the escaping alpha will produce additional scintillation light, allowing to reject this background~\cite{lang2009a}.

\subsection{Light Quenching}\label{sec:quenching}

In a compound material like $\n{CaWO_4}$, neutrons will be seen above threshold mainly if they scatter from light elements (like oxygen) due to simple two-body kinematics. On the other hand, we saw already in the introduction that we can expect WIMPs to scatter mainly from heavy elements (like tungsten) due to the $\propto A^2$ enhancement of the coherent scattering cross section. With scintillators, different recoiling nuclei can be discriminated by their different light yield. This opens the possibility to discriminate the neutron background from WIMP induced events (as required by challenge~\ref{req:neutrons}) even when the neutrons do not double scatter, a feature unique to this method. To this end, precise knowledge of the light yield from tungsten recoils is required. 

We define the light yield of an event as detected light over detected energy, normalized to unity for electrons of $122\1{keV}$. One approach to measure this value uses neutron scattering on a $\n{CaWO_4}$ target at room temperature~\cite{jagemann2005}. The crystal is irradiated with neutrons of known energy from an accelerator, and the scintillation light is measured with a photomultiplier. The angle and the time of flight of the scattered neutrons are recorded. Hence, the kinematics are fixed, and the energy transfer can be calculated. This allows to identify on which nucleus the neutron scattered, and hence to derive the light yield for nuclei present in the crystal. However, for tungsten recoils, only a limit on their light yield could be given~\cite{jagemann2006}. 

The measurement with the usual cryogenic setup with superconducting thermometers for both energy and light determination proves to be difficult, since the observed nuclear recoils induced by common neutron sources are dominantly oxygen recoils, whereas tungsten recoils become dominant only well below $10\1{keV}$. Only very recently this approach was successful in giving a value of the light yield of tungsten recoils using a dedicated cryogenic setup with accelerator produced neutrons~\cite{lanfranchi2009}.

Experiments combining both above variants are under way~\cite{coppi2006}. Measuring the recoil energy via the cryogenic phonon detector as well as the angle of the scattered neutron and its time-of-flight gives redundant information which helps to reduce systematic errors.

A third way to measure the light yield of various recoiling nuclei uses a mass spectrometer technique. The starting point for this approach is the understanding that the scintillation light is not caused by the incident gamma, neutron, or WIMP, but by the recoiling nucleus. Hence, different ions can be accelerated onto the $\n{CaWO_4}$ target. The only difference between such external nuclei and nuclear recoils is due to the binding energy of an atom within the crystal, but since this is of the order of eV, it can safely be neglected when dealing with nuclei which have energies in the keV range. Therefore, in the absence of surface effects, the light output following such an external irradiation will just be the same as if the nucleus recoiled following a particle interaction within the crystal. This method has the major advantage that a large variety of different nuclei can be studied. In addition, the energy of the impinging nucleus can be freely adjusted in the range of interest. Making use of the known time-of-flight of the nuclei through the mass spectrometer allows for clean trigger conditions and hence a low energy threshold. Measurements were done using a $5\times5\times5\1{mm^3}$ cubic $\n{CaWO_4}$ crystal at room temperature, and the scintillation light was read out with a photomultiplier~\cite{ninkovic2006,bavykina2007}.

\begin{figure}[htbp]
\begin{flushright}\includegraphics[angle=90,width=0.8\columnwidth,clip,trim=0 0 130 0]
{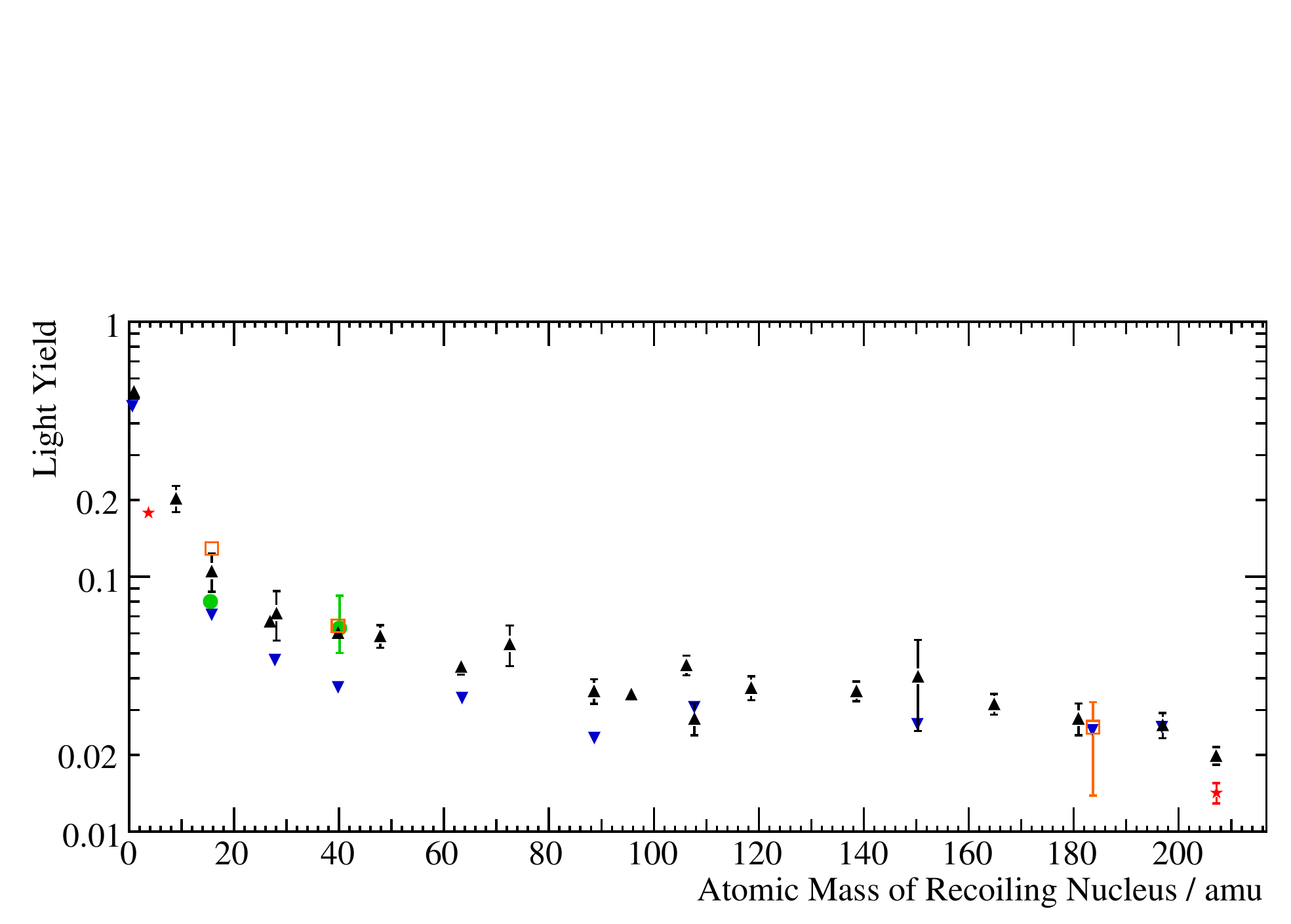}\end{flushright}
\vspace{-5mm}\caption{Various measurements of the light yield of different recoiling nuclei: {\color{blue}{$\blacktriangledown$}}:~From the mass spectrometer technique at room temperature~\cite{ninkovic2006}. {\color{black}{$\blacktriangle$}}:~Refined measurements with the mass spectrometer technique at room temperature, also using a refined error estimation~\cite{huff2006}. {\color{green}{$\bullet$}}:~From neutron scattering at room temperature~\cite{jagemann2004}. {\color{orange}{$\square$}}:~Preliminary results from neutron scattering at cryogenic temperatures~\cite{lanfranchi2009}. {\color{red}{$\bigstar$}}:~In situ measurements at cryogenic temperatures using the CRESST-II setup~\cite{cozzini2004,lang2009a}.}
\label{fig:variousquenchingsF}\end{figure}

Figure~\ref{fig:variousquenchingsF} shows the light yield for various recoiling nuclei, measured at different temperatures and using the above independent methods. Two additional points from in situ measurements with the CRESST-II setup are shown in the graph: One comes from alpha events observed following decays of contaminations in the crystals, giving the light yield of recoiling helium nuclei~\cite{cozzini2004}. Another in situ measurement comes from a background of impinging lead nuclei~\cite{lang2009a}. We note that the error bars given in~\cite{ninkovic2006} are only errors of a fit and may overestimate the precision of the experiment; they are hence not shown in figure~\ref{fig:variousquenchingsF} for clarity. Notable differences between the various measurements only occur for the measurement of the light yield of $\n{{}^{206}Pb}$ at room temperature~\cite{huff2006} and using the CRESST-II setup~\cite{lang2009a}, which is not yet explained. A general trend is observable which can be understood theoretically~\cite{bavykina2007}. To summarize, we can expect the light yield of alpha particles as $0.175\pm0.003$, of oxygen recoils as $0.103\pm0.006$ and of tungsten recoils as $0.025\pm0.002$ (values relative to the light yield of electrons).

\section{The CRESST-II Experiment}

\begin{figure}[htbp]
\begin{flushright}\includegraphics[width=0.8\columnwidth]{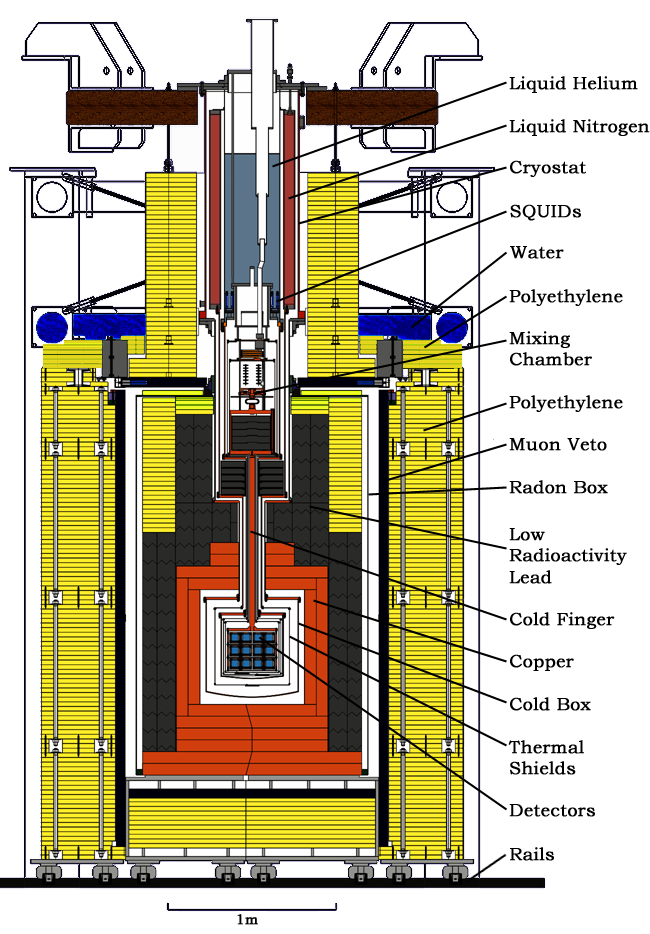}\end{flushright}
\vspace{-5mm}\caption{The CRESST-II setup. The cryostat, in the upper half of the picture, is kept away from the detectors since it is made from standard, i.e. non-radiopure, materials. The shielding consists of copper (orange), lead (grey), the radon box (thin line), the muon veto (blue), the neutron moderator (polyethylene in yellow, water in blue and the helium and nitrogen in the cans of the cryostat). The whole shielding is mounted on wagons movable on rails to allow access to the detectors.}
\label{fig:shielding}\end{figure}

The Cryogenic Rare Event Search with Superconducting Thermometers CRESST-II~\cite{cresst0001} is located in the Laboratori Nazionali del Gran Sasso~\cite{lngs0001} under a minimum rock overburden of 1400m of dolomite rock, where the surface muon flux is reduced by six orders of magnitude to about $2\1{m^{-2}h^{-1}}$~\cite{aglietta1998,arpesella1992}. The $\n{{}^3He/{}^4He}$ dilution refrigerator needed for operation at millikelvin temperatures is commercially available but contains various materials that are not suited for low background applications. This gives the CRESST-II setup its peculiar arrangement, shown in figure~\ref{fig:shielding}. In this setup, the cryostat is kept away from the shielded detectors, and the cooling power is transferred into the shielded volume via a $1.3\1{m}$ long cold finger made from ultra-pure copper. The current $80\1{mK}$ shield of the cryostat encloses $24$ liters which are available as experimental volume. A support structure is installed that can hold up to 33 detector modules, allowing for a target mass of $10\1{kg}$ when equipped with the current $\n{CaWO_4}$ crystals. All parts within the experimental volume are custom made, mostly from ultra pure copper~\cite{angloher2002,angloher2009}. The experimental volume is surrounded, from the inside to the outside, by the thermal shields of the cryostat, $14\1{cm}$ of copper and $20\1{cm}$ of lead against gamma particles, a container that is constantly flushed with nitrogen gas to displace radon, a muon veto, as well as $45\1{cm}$ of polyethylene to shield the ambient neutron flux~\cite{wulandari2004b,wulandari2004}, see figure~\ref{fig:shielding}. A two-storey Faraday cage protects the sensitive electronics, and its lower half is equipped as a clean room. The trigger rate of one crystal operated underground is about $2\1{Hz}$ but drops by two orders of magnitude to about one electron recoil event per minute when the shields are closed, thus meeting the shielding requirement~\ref{req:shielding}.

\section{Detector Operation}~\label{sec:operation}

To facilitate the operation of superconducting transition thermometers for the search for dark matter, an additional heater structure is used. This heater consists of a gold film with a thickness of a few 10s of nanometers that is evaporated directly onto the thermometer film. Pulses of given energies are injected to this gold film in regular time intervals. The response pulse of the thermometer to this injected energy can then be used to infer a variety of very useful informations, as we describe in the following.

\subsection{The Operating Point}

A superconducting phase transition thermometer is characterized by its transition curve. The lower the transition temperature, the smaller the heat capacity of the thermometer film, hence the larger the signal for a given interaction. The steeper the transition, the better the sensitivity of the film. Conversely, the wider the transition, the larger the linear dynamic range of the thermometer. The latter is however not a crucial point since even pulses that drive the thermometer normal conducting can readily be analyzed with still excellent energy resolution, as we will see in section~\ref{sec:w180}.

For a fixed bias current, heater pulses of varying amplitude are injected to the heater structure, thus directly probing the response of the detector to various injected energies. One is then free in choosing the operating point to give the detector the desired properties. The thermometers are usually set up at the top half of the transition since the noise is generally smaller there~\cite{moseley1984,enss2005}.

\subsection{Stability Control}

The low energy interactions of interest here result in temperature changes of the thermometer film of a few $\n{\mu K}$. Hence, the film temperature needs to be stabilized to similar precision. This requires a weak thermal coupling of the detector modules to the cryostat as well as an active temperature control. To this end, heater pulses are injected to the heater structure every few seconds, which is frequent enough to sample temperature variations of the cryostat, but does not introduce too much dead time. These injected heater pulses are large enough to drive the thermometer out of its transition. The response pulse from the thermometer is evaluated online and serves as an input variable for a PI control of the operating point through an adjustment of a steady current through the heater structure. 

\begin{figure}[htbp]
\begin{flushright}\includegraphics[width=0.8\columnwidth]{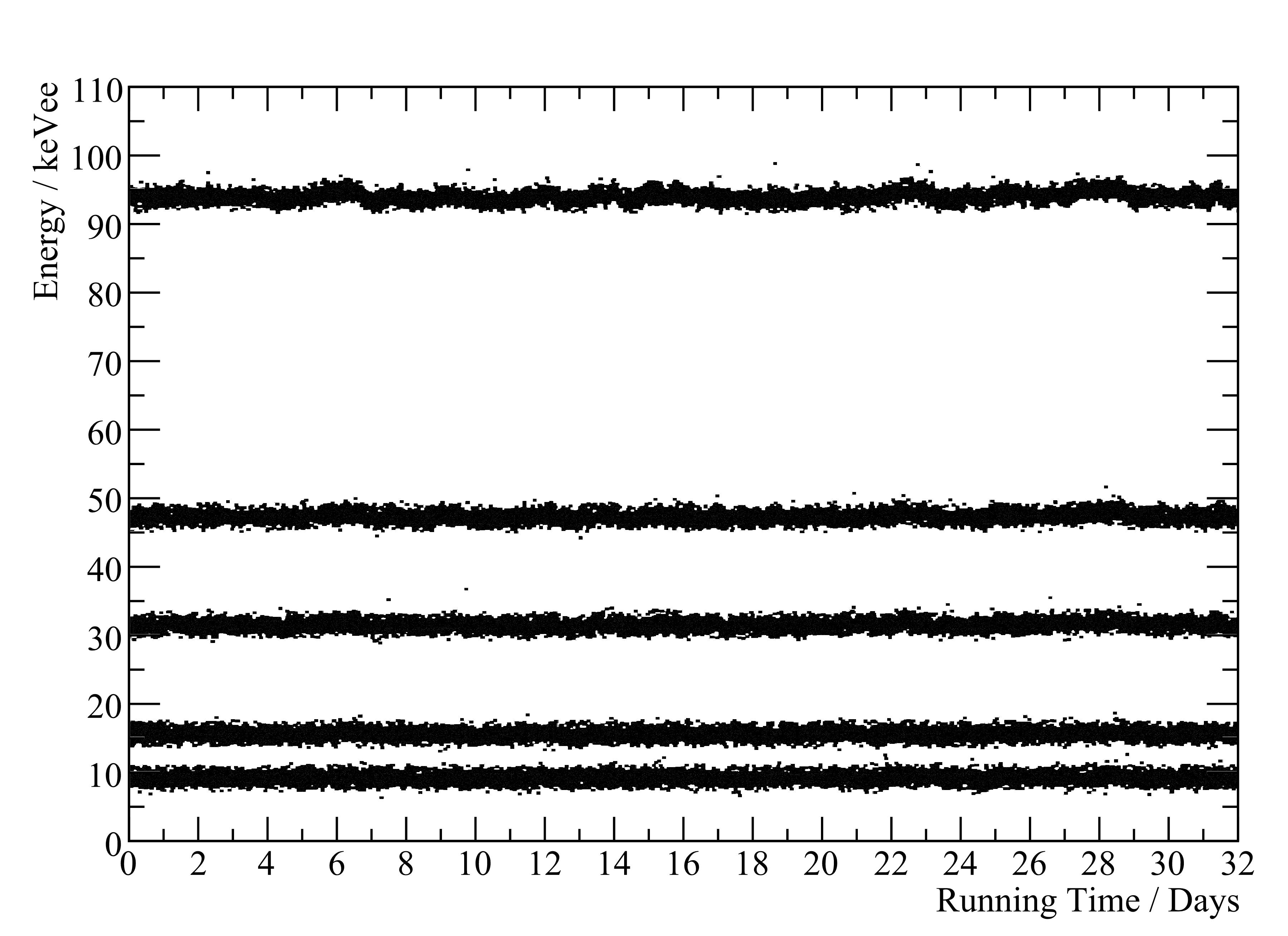}\end{flushright}
\vspace{-5mm}\caption{The amplitude response of the light detector to injected heater pulses is constant within resolution and demonstrates the high stability of the detectors over many weeks of running time~\cite{angloher2005}. The energy is in units of $\n{keV_{ee}}$ (keV electron equivalent), defined such that an electron of $122\1{keV}$ deposits $122\1{keV_{ee}}$ in the light detector.}
\label{fig:stability}\end{figure}

Figure~\ref{fig:stability} shows the amplitude response of the light detector to injected heater pulses, demonstrating the highly stable conditions achievable. The small width of observed spectral lines in the phonon detector (see section~\ref{sec:spectralfeatures}) is only obtainable under such stable running conditions. In addition, the readiness of the detector to identify dark matter interactions is monitored by this procedure, hence meeting challenge~\ref{req:long} concerning long measuring times.

\subsection{Energy Calibration}

\begin{figure}[htbp]
\begin{flushright}\includegraphics[angle=90,width=0.8\columnwidth]{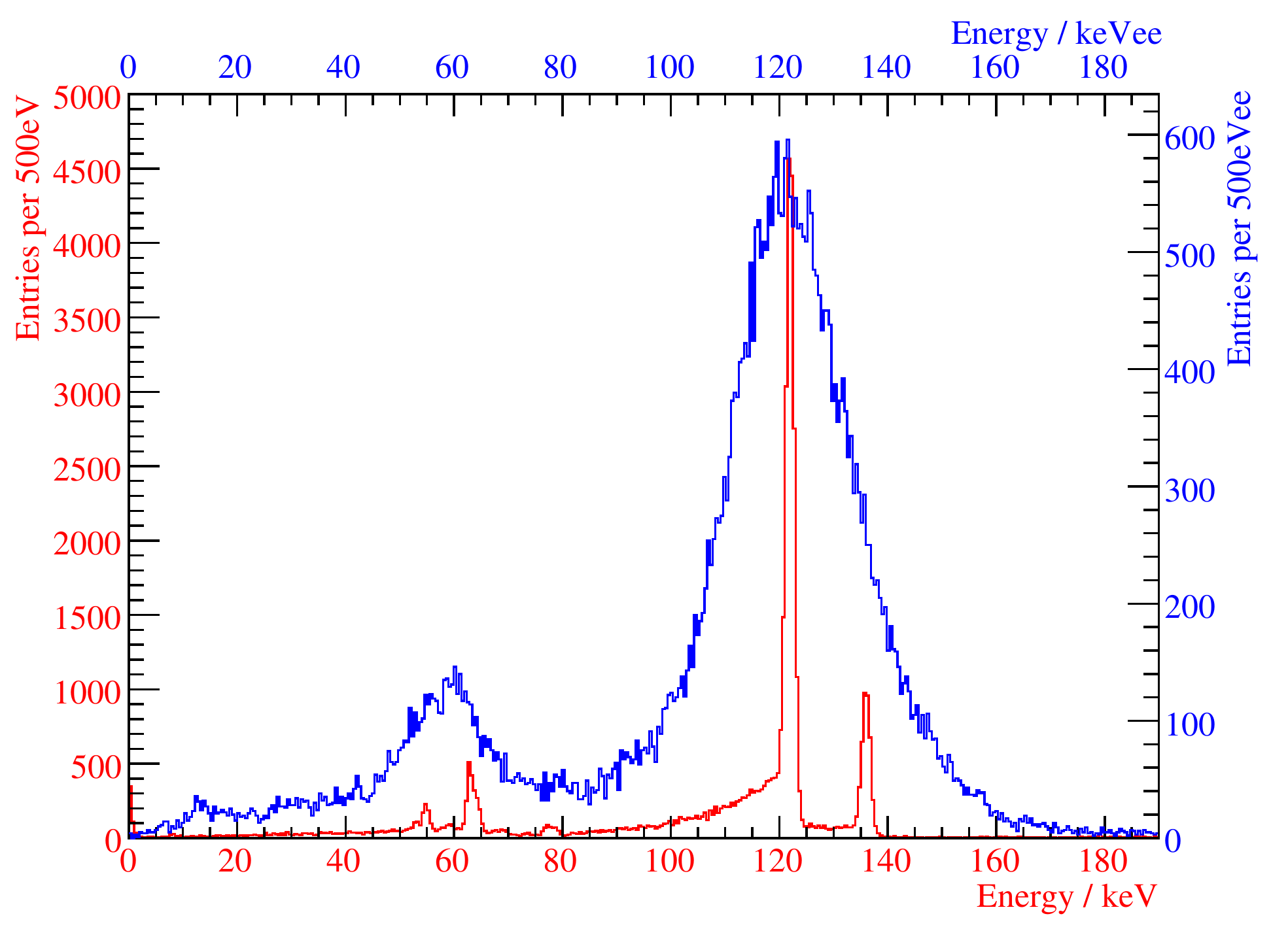}\end{flushright}
\vspace{-5mm}\caption{Data from a $\n{{}^{57}Co}$ calibration of a phonon detector (lower spectrum in red and scales on the left and bottom) and its light detector (upper spectrum and scales on the top and right). The resolution of the phonon channel is an order of magnitude better than that of the light channel. The $122\1{keV}$ and $136\1{keV}$ lines from $\n{{}^{57}Co}$ and escape lines between $(50-80)\1{keV}$ are visible.}
\label{fig:OverlaidSpectraE}\end{figure}

A $\n{{}^{57}Co}$ calibration gives calibration lines at $122\1{keV}$ and $136\1{keV}$ as well as a set of escape lines between $(50-80)\1{keV}$. Figure~\ref{fig:OverlaidSpectraE} shows a typical calibration spectrum of a phonon detector and its light channel~\cite{lang2009b}.

The targets for cryogenic scintillator searches are enclosed in a cryostat, which in CRESST-II constitutes a minimum thickness of $12\1{mm}$ of copper around the crystals. This barrier is only penetrated by $\order(100\1{keV})$ gammas, but the energy range of interest is $\order(10\1{keV})$. To meet the energy calibration challenge~\ref{req:calibration}, a method is therefore required to take the energy calibration down to lower energies. This is also achieved with the help of the heater structure on the thermometer. Heat pulses with energies in the range of interest are injected to the heater at regular time intervals, typically every 30~seconds. Since one has the injected energy under control, the amplitude response of the thermometer can then be used to take the energy calibration from $\order(100\1{keV})$ down to below $10\1{keV}$~\cite{lang2009b}.

This energy calibration procedure is validated by the location of spectral lines at the expected energies. A variety of such features can be identified. Of particular interest are spectral features recorded during the search for dark matter, which constitute an in situ calibration during the long exposure~\cite{lang2009b}. The lowest energy example of such a feature is the decay of $\n{{}^{41}Ca}$ at $3.6\1{keV}$ (see section~\ref{sec:spectralfeatures}), satisfying the energy calibration challenge~\ref{req:calibration}.

\subsection{Trigger Efficiency}

\begin{figure}[htbp]
\begin{flushright}\includegraphics[angle=90,width=0.8\columnwidth,clip,trim=0 0 155 0]{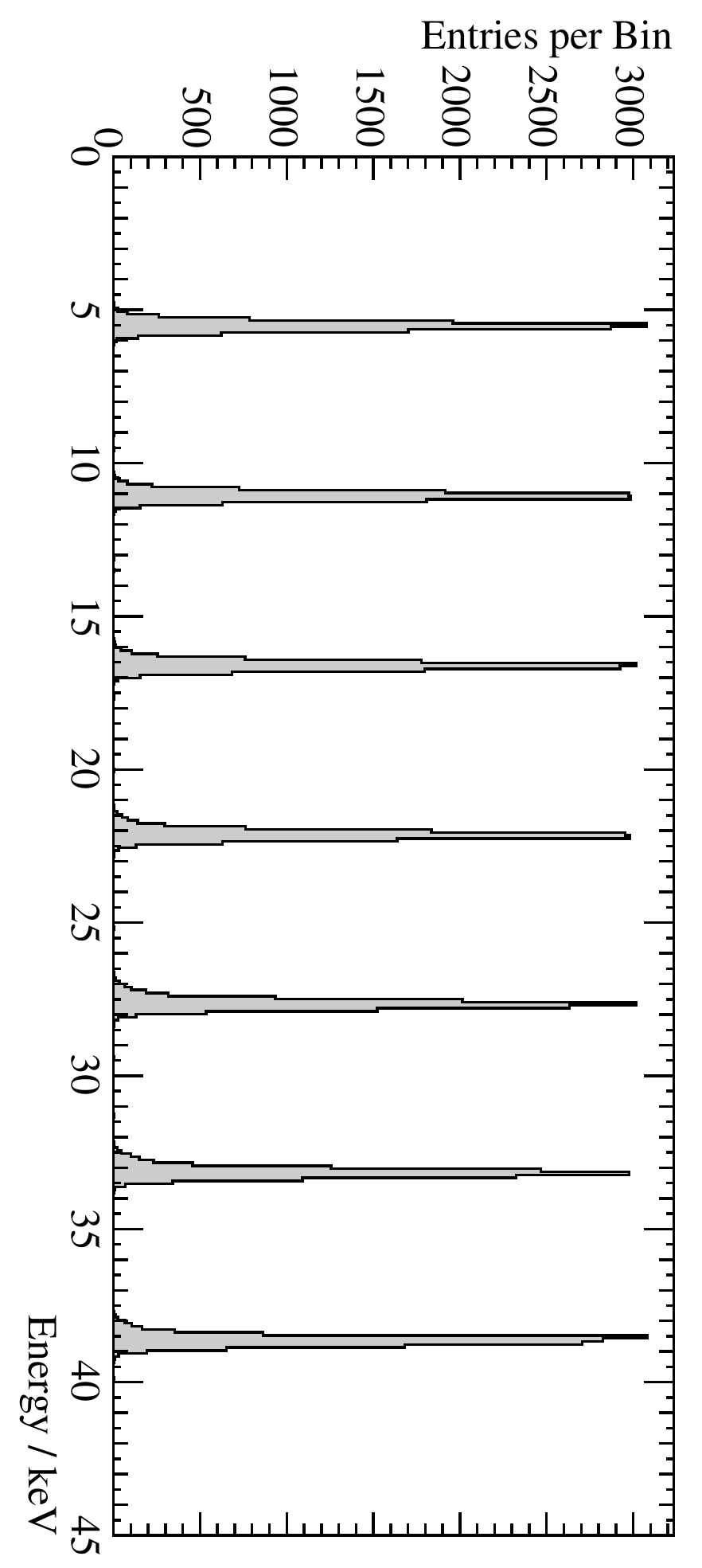}\end{flushright}
\vspace{-5mm}\caption{Heater pulses from a set of fixed energies are injected every few seconds throughout the WIMP search. The number of triggered responses is shown here and can be seen to be constant, proving the constant trigger efficiency in the probed energy range. For the search for dark matter, an analysis threshold of $10\1{keV}$ is imposed~\cite{angloher2009}, well in the range of constant trigger efficiency.}
\label{fig:Trigger}\end{figure}

The trigger efficiency to low energy nuclear recoils is a crucial parameter especially when a limit on the WIMP-nuclear scattering cross section is to be placed. Any experiment aiming to do so needs to demonstrate its trigger efficiency over the full energy range used for the WIMP analysis, as pronounced in requirement~\ref{req:trigger}. In addition to the usual usage of a neutron calibration and its comparison to expected scatter rates, the heater structures on CRESST-II detectors provide the possibility to directly measure this value. Heat pulses are injected to the phonon detector over the full energy range of interest. The number of pulses that trigger are a direct measure of the trigger efficiency, and can be seen in figure~\ref{fig:Trigger} to be constant over the full energy range probed.

\subsection{Neutron Calibration}

\begin{figure}[htbp]
\begin{flushright}\includegraphics[angle=90,width=0.8\columnwidth,clip,trim=0 0 25 0]{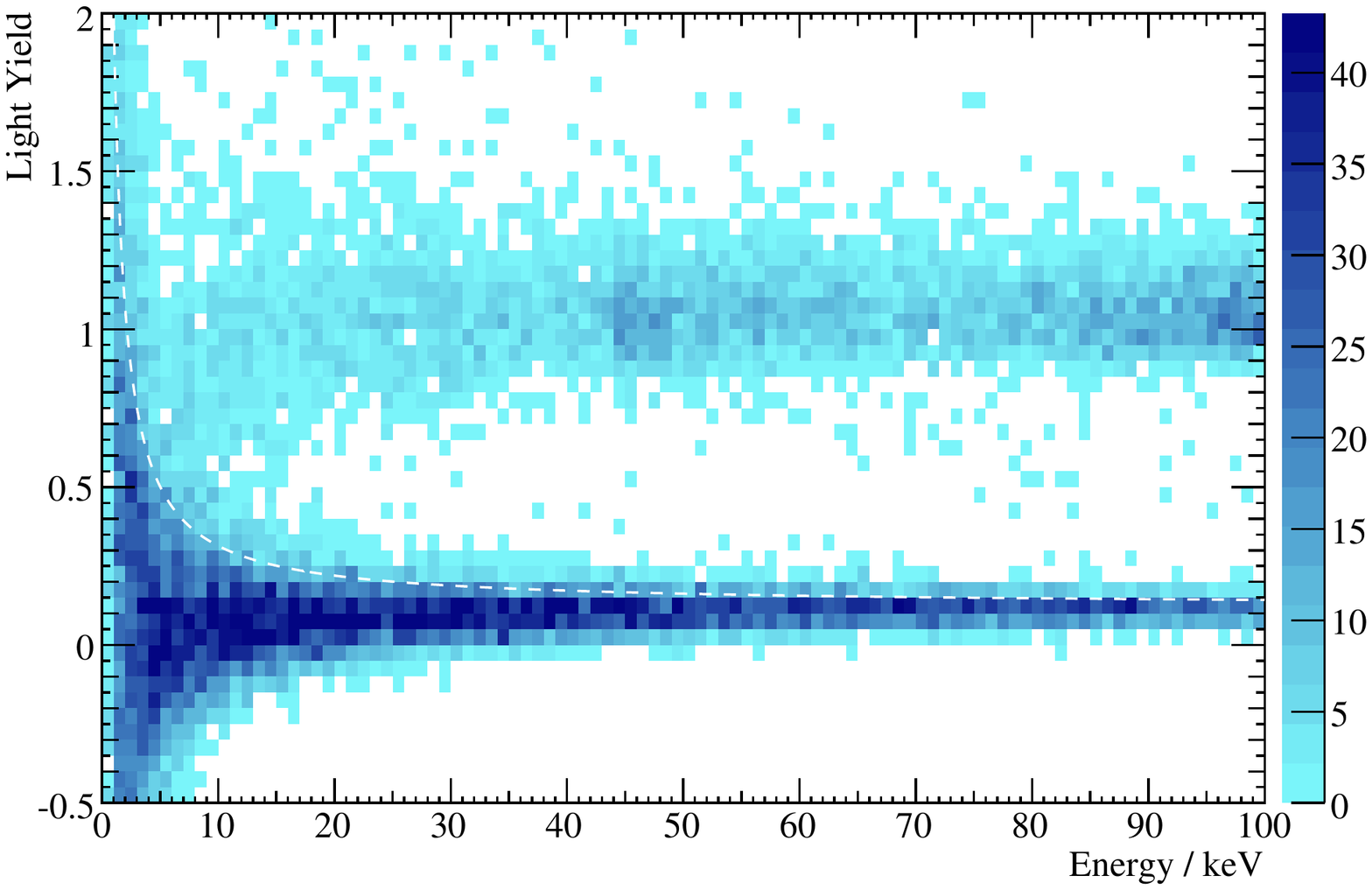}\end{flushright}
\vspace{-5mm}\caption{Testing a detector with an external neutron source. Negative light yield values arise from the fitting procedure which allows for negative amplitudes in order to treat baseline noise in an unbiased way. The upper electron recoil band as well as the lower nuclear recoil band are well separable above $\sim10\1{keV}$. Below the dashed line, $90\percent$ of oxygen recoil events are expected based on the independent light yield measurements described in section~\ref{sec:quenching}. The data can be seen to agree well with this expectation~\cite{angloher2009}.}
\label{fig:VerenaNCalxc}\end{figure}

The position of the gamma and electron recoil band and the resolution of the light detector is determined from calibration data or directly from the WIMP search data. The position of the expected WIMP induced signal region is then calculated based on the light yield of tungsten recoils. To demonstrate the validity of this approach, data is taken with a neutron source present. The position of the neutron induced nuclear recoil band is expected to follow the calculated band of oxygen recoils. This can indeed be seen in figure~\ref{fig:VerenaNCalxc}, where the calculated line below which $90\percent$ of oxygen recoils are expected is also drawn, hence validating this procedure~\cite{angloher2009}.

\section{Results}

The main objective of the CRESST experiment is of course the search for dark matter. Yet the unique properties of the employed detectors led to a variety of other notable results. In this section, we briefly review these results before presenting the current status of the experiment regarding the search for dark matter.

\subsection{Fracture Processes}\label{sec:cracks}

First measurements using large (non-scintillating) sapphire crystals led to a trigger rate orders of magnitude above the expected level. A statistical analysis showed that events were not Poissonian distributed. Eventually these events were identified as being due to cracks developing in the crystals due to a too tight clamping~\cite{astrom2006a,astrom2006b}. Today, such devices can be used to make the most sensitive measurements of stress relaxation in solids~\cite{astrom2007}, and even to learn about earthquakes~\cite{astrom2006c}. For the search for dark matter, cracking was eliminated by the use of soft clamps~\cite{majorovits2009}.

\subsection{Tungsten Decay}\label{sec:w180}

\begin{figure}[htbp]
\begin{flushright}\includegraphics[angle=90,width=0.8\columnwidth,clip,trim=0 0 155 0]{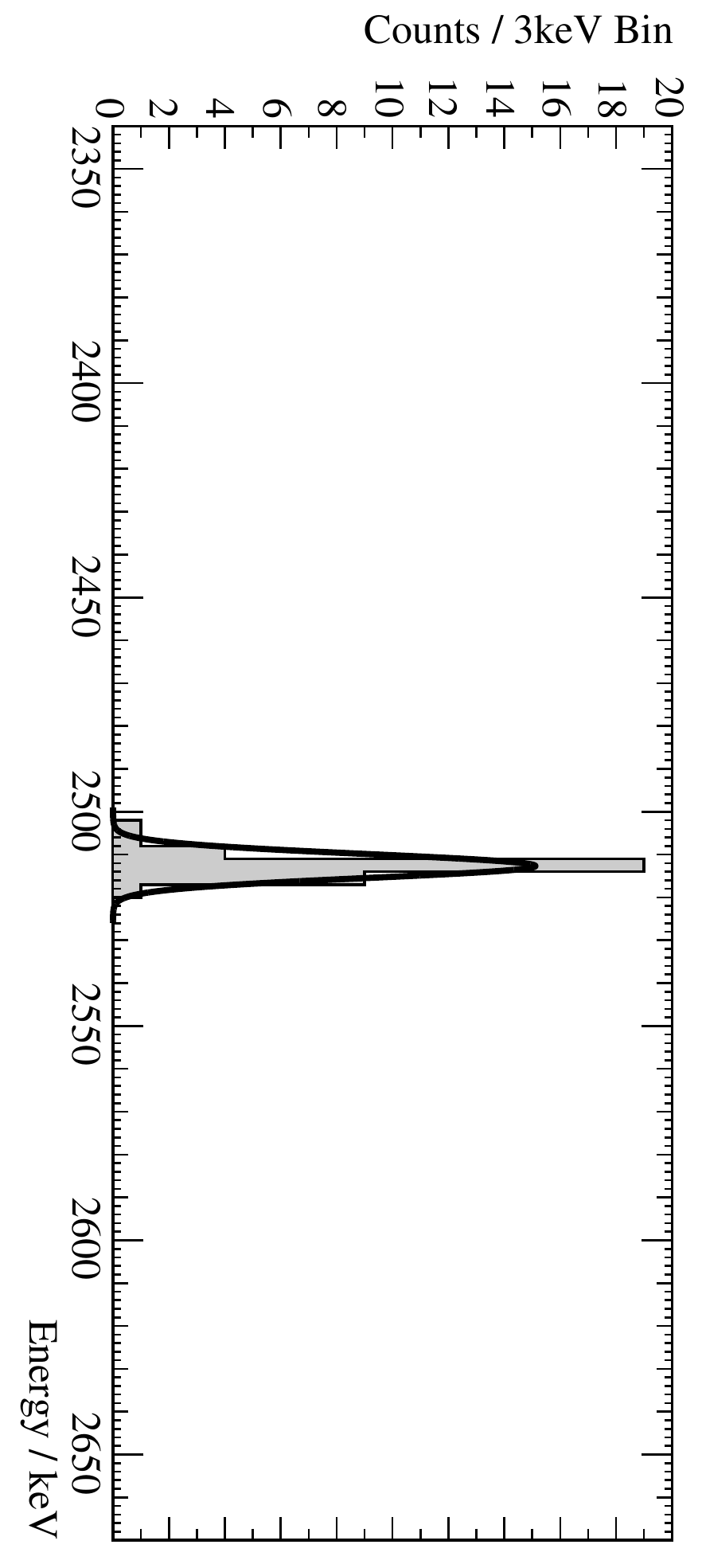}\end{flushright}
\vspace{-5mm}\caption{Alpha events observed in one crystal with an exposure of $12.3\1{kg\,d}$ in a wide energy interval around the peak from the $\n{{}^{180}W}$ alpha decay, together with a Gaussian fit to the peak. The measurement can be seen to be basically background free~\cite{cozzini2004} and constitutes the first unambiguous measurement of this decay.}
\label{fig:180wb}\end{figure}

All naturally occurring tungsten isotopes are expected to alpha decay into hafnium, but with extremely long lifetimes. Since the decay energies for all these decays are in the same energy range as beta and gamma backgrounds from the natural decay chains, their observation is a difficult task. Yet with cryogenic scintillator experiments, these backgrounds can be discriminated from the alpha signal, leading to a basically background free measurement of such alpha decays, see figure~\ref{fig:180wb}. Hence, the natural decay of $\n{{}^{180}W}$ was observed unambiguously for the first time. The half-life was determined to be $T_{1/2}=(1.8\pm0.2)\times10^{18}\1{years}$ at a precisely determined Q-value of  $(2516.4\pm1.1(\n{stat})\pm1.2(\n{sys}))\1{keV}$~\cite{cozzini2004}, consistent with an earlier indication~\cite{danevich2003}. Also, most stringent limits on the half-lives of the other natural tungsten isotopes of the order of $T_{1/2}>8\times10^{21}\1{years}$ could be set.

\subsection{Cosmogenic Activation}\label{sec:spectralfeatures}

\begin{figure}[htbp]
\begin{flushright}\includegraphics[angle=90,width=0.8\columnwidth,clip,trim=0 50 132 0]{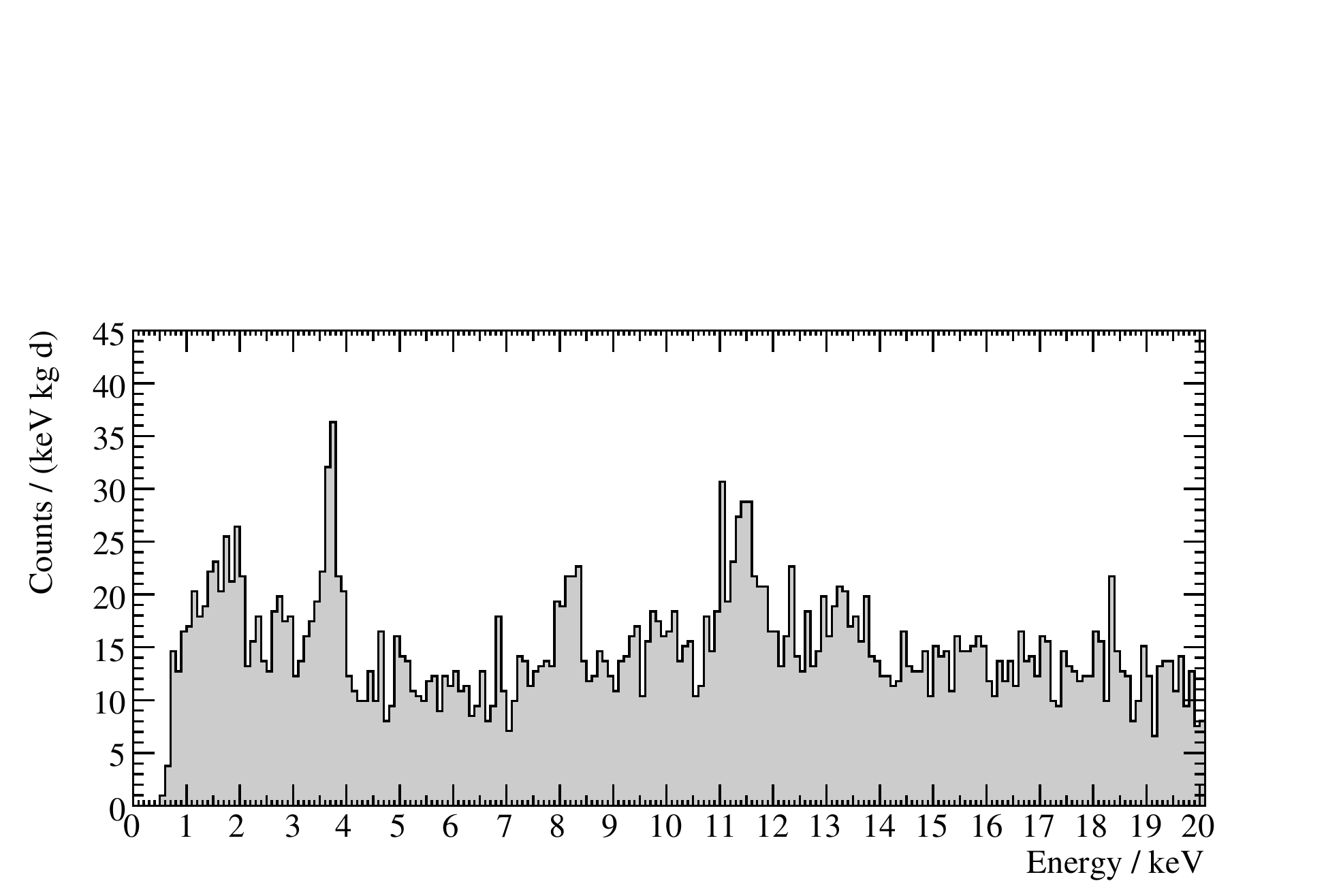}\end{flushright}
\vspace{-5mm}
\caption{A low energy background spectrum, recorded with one crystal after an exposure of $21.2\1{kg\,d}$~\cite{angloher2009}. The line from  $\n{{}^{41}Ca}$ at $3.61\1{keV}$ is clearly visible~\cite{lang2009b}. The line at $8.0\1{keV}$ is due to copper fluorescence from the surrounding cryostat.}
\label{fig:verenarun30l}\end{figure}

All materials that are exposed to cosmic rays are activated. The activation of tungsten in reactions $\n{{}^{182}W} (p,\alpha) \n{{}^{179}Ta}$ or $\n{{}^{183}W} (p,t) \n{{}^{181}W}$ leads to X-ray lines at $65.4\1{keV}$ and $73.7\1{keV}$, respectively~\cite{mortensen1980,khandaker2008,firestone1996}, which are indeed observed in the spectra taken during WIMP searches~\cite{lang2009b,majorovits2006}. The cosmic activation of calcium results in $\n{{}^{41}Ca}$, which decays with a half-life of $10^5$ years via electron-capture to $\n{{}^{41}K}$. The resulting signal in the calorimeter is the K shell binding energy of potassium at $3.61\1{keV}$. This line has been observed with the CRESST-II detectors with an activity of only $(26\pm4)\1{\mu Bq}$~\cite{lang2009b}, corresponding to only one $\n{{}^{41}Ca}$ isotope in $(2.2\pm0.3)\times10^{16}$ $\n{{}^{40}Ca}$ atoms, see figure~\ref{fig:verenarun30l}. This constitutes the most precise measurement of this abundance to date, more than an order of magnitude more sensitive than other measurements which are typically done by accelerator mass spectrometry~\cite{merchel2009,fink1990}.

\subsection{WIMP Search}

\begin{figure}[htbp]
\begin{flushright}\includegraphics[angle=90,width=0.8\columnwidth,clip,trim=0 0 25 0]{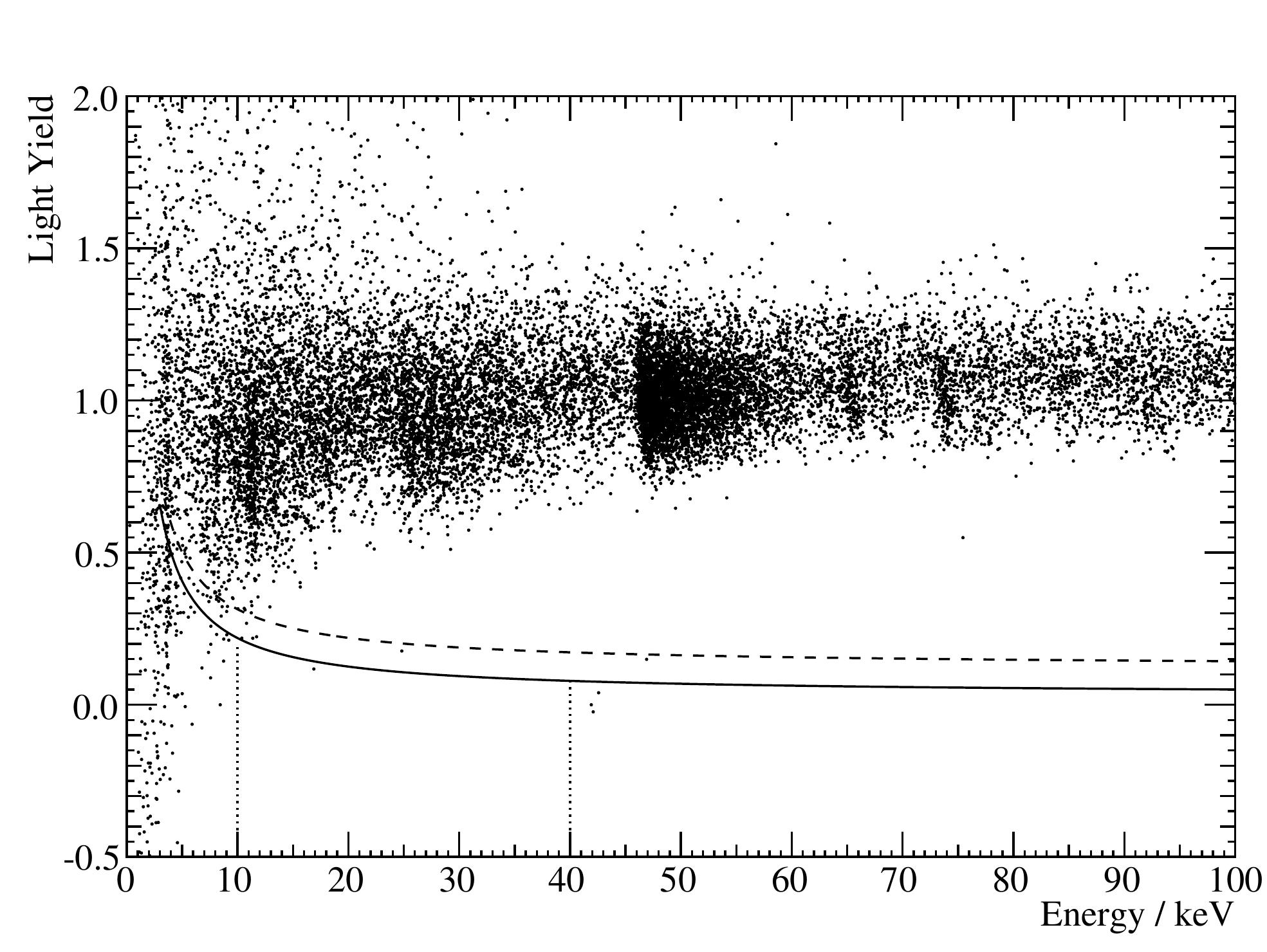}\end{flushright}
\vspace{-5mm}\caption{Scatter plot of events observed in one detector with an exposure of $24.1\1{kg\,d}$~\cite{angloher2009}. Below the dashed line, $90\percent$ of all (mostly neutron induced) oxygen recoils are expected. Below the solid line, $90\percent$ of all (possibly WIMP induced) tungsten recoils are expected. The acceptance region in the WIMP search is bounded by this latter curve, and extends in energy between $10\1{keV}$ (where background discrimination becomes efficient) and $40\1{keV}$ (above which form factor effects suppress the signal). The origins of the observed spectral features in the electron band are explained in~\cite{lang2009b}.}
\label{fig:VerenaRun30x}
\end{figure}

\begin{figure}[htbp]
\begin{flushright}\includegraphics[angle=90,width=0.8\columnwidth,clip,trim=0 0 25 0]{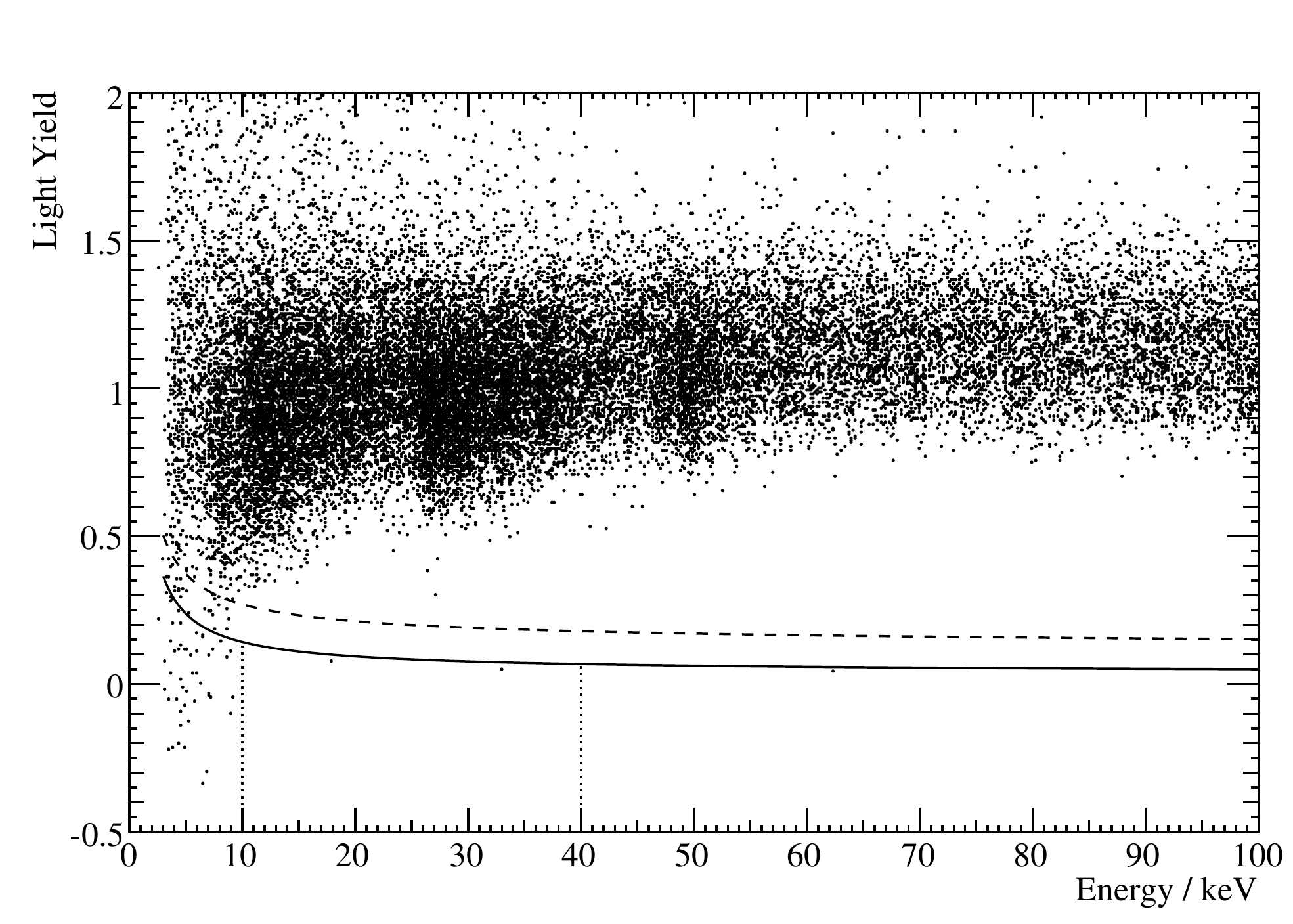}\end{flushright}
\vspace{-5mm}\caption{Scatter plot of events observed in another detector with a similar exposure of $23.8\1{kg\,d}$~\cite{angloher2009}, compare with figure~\ref{fig:VerenaRun30x}.}
\label{fig:ZoraRun30x}\end{figure}

The coherent WIMP-nucleon scattering cross section has been constrained with data from two detector modules taken during the prototyping phase of the CRESST-II experiment~\cite{angloher2005}, and from two detector modules taken during the commissioning phase of the new setup~\cite{angloher2009}. Data from the latter are shown in figures~\ref{fig:VerenaRun30x} and~\ref{fig:ZoraRun30x} after an exposure of about three months to background radiation alone.

\begin{figure}[htbp]
\begin{flushright}\includegraphics[angle=90,width=0.8\columnwidth,clip,trim=0 0 100 0]{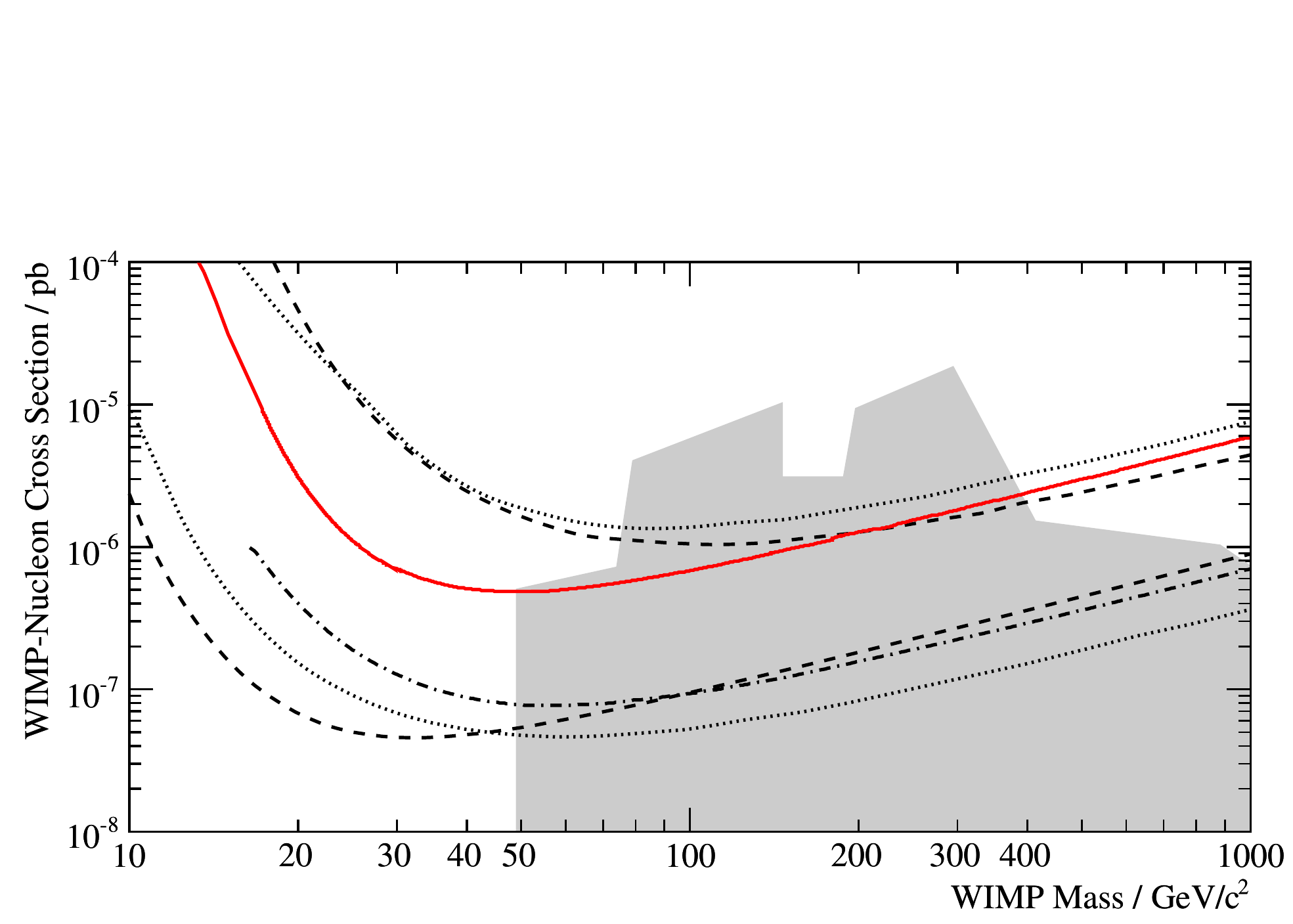}\end{flushright}
\vspace{-5mm}\caption{Current upper limits on the coherent WIMP-nucleon cross section as a function of WIMP mass. The solid line comes from the CRESST-II experiment~\cite{angloher2009}. The grey region is a prediction from theory~\cite{baltz2003}. Limits from other experiments are also shown: CDMS~\cite{ahmed2009} (lower dotted) and EDELWEISS~\cite{sanglard2005} (upper dotted), XENON10~\cite{angle2008} (lower dashed) and WARP~\cite{benetti2008} (upper dashed), and ZEPLIN-III~\cite{lebedenko2009} (dash-dotted).}
\label{fig:limits}\end{figure}

In the absence of a clear signal, a limit on the coherent WIMP-nucleus scattering cross section is calculated using standard assumptions on the dark matter halo~\cite{donato1998,lewin1996}. The fact that a nucleus is not point-like is taken into account by assuming the Helm form factor~\cite{helm1956}, which basically limits the energy transfer to the tungsten nuclei to energies below $40\1{keV}$ for all WIMP masses. In the energy region above $10\1{keV}$, where recoil discrimination becomes efficient, to up to $40\1{keV}$, three events in the tungsten recoil area were observed in the data of figures~\ref{fig:VerenaRun30x} and~\ref{fig:ZoraRun30x}. The upper limit for the WIMP scattering cross-section per nucleon is set using Yellin's maximum gap method~\cite{yellin2002}, and shown as the solid curve in figure~\ref{fig:limits}. The minimum of this curve is below $5\times10^{-7}\1{pb}$.

\begin{figure}[htbp]
\begin{flushright}\includegraphics[angle=90,width=0.8\columnwidth,clip,trim=0 0 80 0]{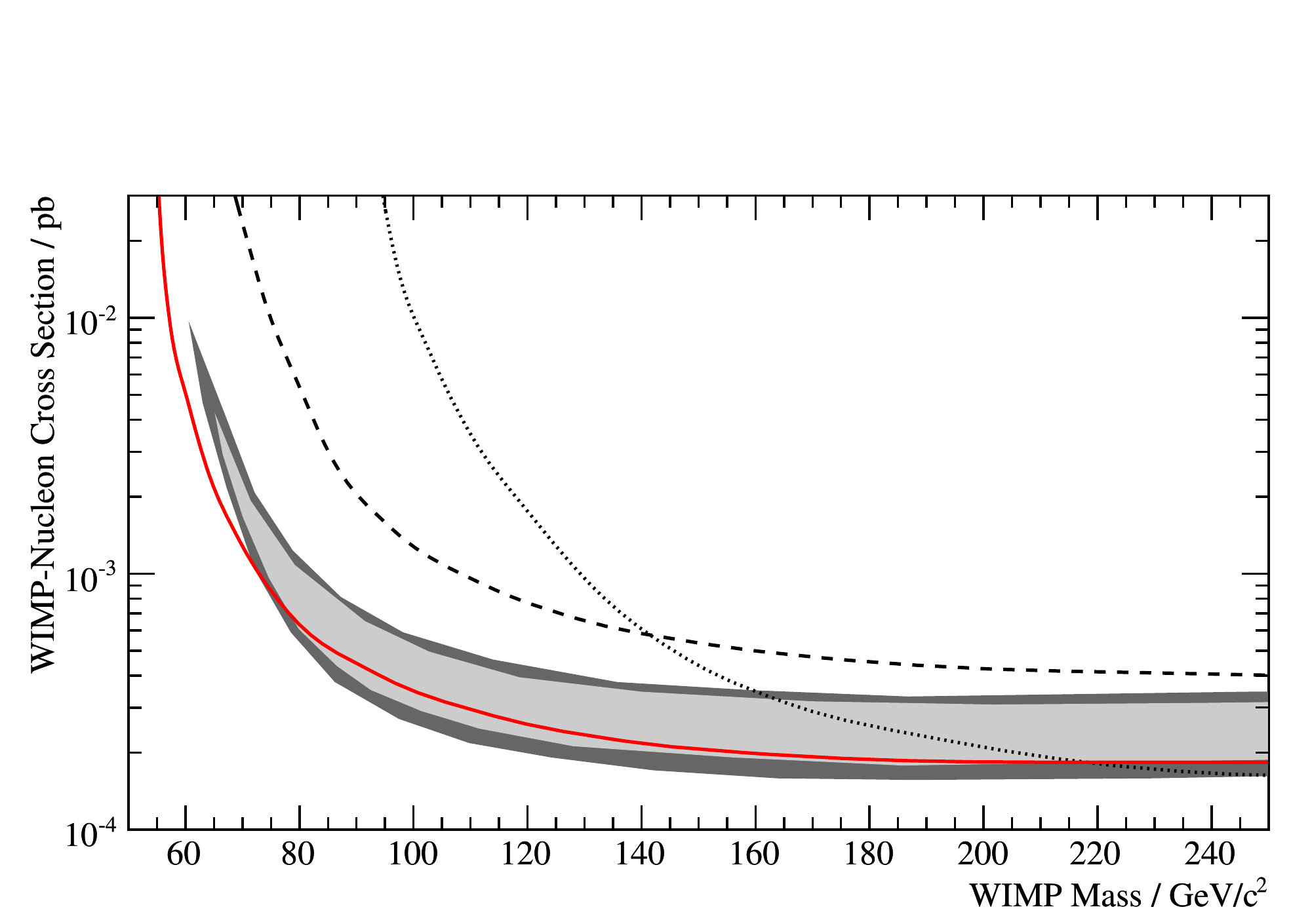}\end{flushright}
\vspace{-5mm}\caption{Current upper limits on the WIMP-nucleon scattering cross section from the CRESST-II experiment (red solid line) as a function of WIMP mass in an Inelastic dark matter model with an excited WIMP state $120\1{keV}$ above the ground state, and for a Galactic escape speed of $500\1{km/s}$~\cite{chang2008}. The shaded areas are the $90\percent$ and $99\percent$ confidence contours from the DAMA/NaI and DAMA/LIBRA experiment. The constraints from the two most stringent experiments in the standard scenario (figure~\ref{fig:limits}) are also shown for this model: XENON10 (dashed) and CDMS (dotted). More details as well as limits from other experiments can be found in~\cite{chang2008}.}
\label{fig:chang2008d}\end{figure}

Besides the standard WIMP scenario, other models of the nature of dark matter and its interactions have to be considered. One such example is the Inelastic dark matter model~\cite{smith2001,chang2008}. There, the WIMP undergoes a transition to an excited state in the scattering process. Hence, the total energy $E_{\n{CM}}$ that is available in the scattering process needs to be larger than the splitting energy of the excited level, which is $\order(100\1{keV})$ above the ground state. Trivially,
\begin{eqnarray}
	E_{\n{CM}}=\frac{1}{2}\mu v^2
\end{eqnarray}
with the relative velocity of dark matter particle $v$ and the reduced mass $\mu$. Interactions in a heavy target have a large reduced mass $\mu$ and hence a large energy $E_{\n{CM}}$ available to excite internal states of the WIMP. Thus, the CRESST-II experiment, with tungsten as the heaviest target nucleus used in any direct WIMP search today, places the most stringent constraints on these models. Figure~\ref{fig:chang2008d} shows as an example one such exclusion for a particular set of splitting energy and halo parameters. In particular, the figure also shows the exclusion curves from the XENON10 and CDMS experiments for comparison with figure~\ref{fig:limits}.

\section{Conclusions}

We have formulated a set of requirements for any experiment that aims to directly detect the interaction of WIMPs with a target, and shown in turn how these requirements are met by the CRESST-II experiment. We gave examples for the sensitivity of experiments with cryogenic scintillators to rare decays, such as the alpha decay of $\n{{}^{180}W}$ and the ratio $\n{{}^{41}Ca/{}^{40}Ca}$. Data from the search for dark matter were presented together with limits on the scattering cross sections for common WIMPs or in Inelastic dark matter models. 

At the time of writing, more than a dozen of $\n{CaWO_4}$ and $\n{ZnWO_4}$ crystals are installed in the CRESST-II experiment. This allows to improve the current limit by two orders of magnitude in the near future, after an exposure of $\sim1000\1{kg\,d}$, if no signal is observed. Beyond this, the CRESST-II and ROSEBUD collaborations have joined with the EDELWEISS collaboration and other groups to build the ton-scale EURECA detector, utilizing this technology for another increase in sensitivity of two orders of magnitude~\cite{kraus2007b}. 

We thank the editors for the invitation to this contribution, Karoline Sch\"affner for the measurement used to make figure~\ref{fig:transition} and Jens Schmaler for picture~\ref{fig:modulfotok}. We are grateful to Dieter Hauff, Ethan Brown and the referee for numerous useful suggestions that improved the manuscript.

\bigskip

\providecommand{\bysame}{\leavevmode\hbox to3em{\hrulefill}\thinspace}
\providecommand{\MR}{\relax\ifhmode\unskip\space\fi MR }
\providecommand{\MRhref}[2]{%
  \href{http://www.ams.org/mathscinet-getitem?mr=#1}{#2}
}
\providecommand{\href}[2]{#2}

\end{document}